\documentclass[sigconf]{acmart}
\usepackage{xcolor}
\usepackage{enumerate}
\usepackage[utf8]{inputenc} 
\usepackage[T1]{fontenc}    
\usepackage{url}            
\usepackage{booktabs}       
\usepackage{amsfonts}       
\usepackage{nicefrac}       
\usepackage{microtype}      
\usepackage{makecell, cellspace}         
\usepackage{color, colortbl}
\usepackage{graphicx}
\usepackage{bbm}
\usepackage{amsthm}
\usepackage{algorithm}
\usepackage{algpseudocode}
\usepackage{multirow}
\usepackage{adjustbox}
\usepackage{enumitem}
\usepackage{bm}
\usepackage{mathtools}
\usepackage{mwe}
\usepackage{tabularx}
\usepackage{wrapfig}
\usepackage{capt-of}
\usepackage{ulem}

\newcommand{\proposed}{\textsf{MUSE}}

\AtBeginDocument{%
  \providecommand\BibTeX{{%
    \normalfont B\kern-0.5em{\scshape i\kern-0.25em b}\kern-0.8em\TeX}}}

\copyrightyear{2023}
\acmYear{2023}
\setcopyright{acmlicensed}
\acmConference[CIKM '23] {Proceedings of the 32nd ACM International Conference on Information and Knowledge Management}{October 21--25, 2023}{Birmingham, United Kingdom.}
\acmBooktitle{Proceedings of the 32nd ACM International Conference on Information and Knowledge Management (CIKM '23), October 21--25, 2023, Birmingham, United Kingdom}
\acmPrice{15.00}
\acmISBN{979-8-4007-0124-5/23/10}
\acmDOI{10.1145/3583780.3614976}

\begin{document}

\title{MUSE: Music Recommender System with Shuffle Play Recommendation Enhancement}

\author{Yunhak Oh}
\authornote{Both authors contributed equally to this research.}
\orcid{}
\affiliation{%
  \institution{KAIST}
  \city{Daejeon}
  \state{}
  \country{Republic of Korea}
  \postcode{34141}
}
\email{yunhak.oh@kaist.ac.kr}

\author{Sukwon Yun}
\orcid{}
\authornotemark[1]
\affiliation{%
  \institution{KAIST}
  \city{Daejeon}
  \state{}
  \country{Republic of Korea}
  \postcode{34141}
}
\email{swyun@kaist.ac.kr}

\author{Dongmin Hyun}
\orcid{}
\affiliation{%
  \institution{POSTECH}
  \city{Pohang}
  \state{}
  \country{Republic of Korea}
  \postcode{37673}
}
\email{dm.hyun@postech.ac.kr}

\author{Sein Kim}
\orcid{}
\affiliation{%
  \institution{KAIST}
  \city{Daejeon}
  \state{}
  \country{Republic of Korea}
  \postcode{34141}
}
\email{rlatpdlsgns@kaist.ac.kr}

\author{Chanyoung Park}
\authornote{Corresponding author.}
\orcid{}
\affiliation{%
  \institution{KAIST}
  \city{Daejeon}
  \state{}
  \country{Republic of Korea}
  \postcode{34141}
}
\email{cy.park@kaist.ac.kr}


\renewcommand{\shortauthors}{Yunhak Oh, Sukwon Yun, Dongmin Hyun, Sein Kim, \& Chanyoung Park}

\begin{abstract}
    Recommender systems have become indispensable in music streaming services, enhancing user experiences by personalizing playlists and facilitating the serendipitous discovery of new music. However, the existing recommender systems overlook the unique challenges inherent in the music domain, specifically shuffle play, which provides subsequent tracks in a random sequence. Based on our observation that the shuffle play sessions hinder the overall training process of music recommender systems mainly due to the high unique transition rates of shuffle play sessions, we propose a \textbf{Mu}sic Recommender System with \textbf{S}huffle Play Recommendation \textbf{E}nhancement (\proposed).
    \proposed~employs the self-supervised learning framework that maximizes the agreement between the original session and the augmented session, which is augmented by our novel session augmentation method, called transition-based augmentation. To further facilitate the alignment of the representations between the two views, we devise two fine-grained matching strategies, i.e., item- and similarity-based matching strategies.
    Through rigorous experiments conducted across diverse environments, we demonstrate ~\proposed's efficacy over 12 baseline models on a large-scale Music Streaming Sessions Dataset (MSSD) from Spotify. The source code of~\proposed~is available at \url{https://github.com/yunhak0/MUSE}.
\end{abstract}



\maketitle

\section{Introduction}
Recommender systems \cite{gru4rec, narm, stamp, sasrec, srgnn, csrm, cl4srec} play a crucial role in providing an immersive user experience that navigates users to access various online content. Specifically, the recent prominence of Session-based Recommendation (SBR) lies in its ability to leverage implicit feedback gathered during a user's session, i.e., activities within a specified period. Due to their ability to adeptly handle session information, these applications have permeated our daily lives, with examples found across a range of domains, from books \cite{book1, book2}, apparel \cite{fashion1, fashion2}, and movies \cite{movie1, movie2, movie3}.

\begin{figure}[!t]
    \includegraphics[width=1\columnwidth]{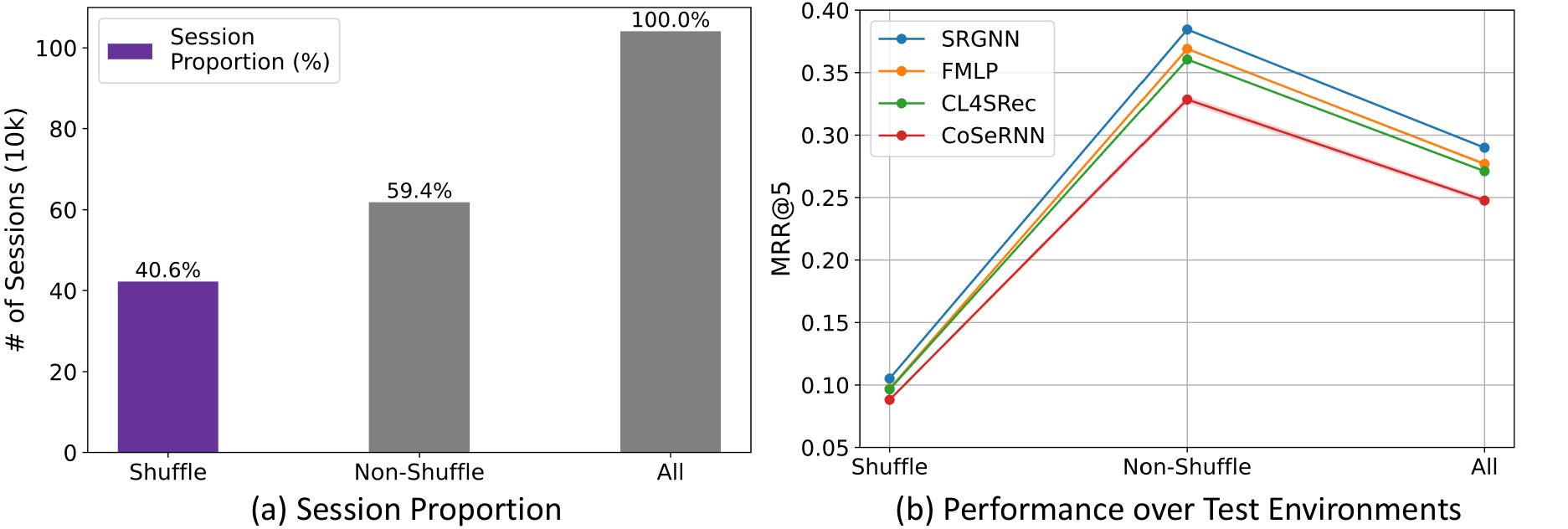}
    \vspace{-5ex}
    \caption[Degradation of shuffle performance]{\label{fig:degrade_shuffle} Recommendation performance (MRR@5) of SBR models and music recommender model on MSSD-5d dataset.}
    \vspace{-4.5ex}
\end{figure}

In contrast to such widely researched domains (e.g., books, fashion, or movies), building a successful music recommender system is especially challenging due to the inherent characteristics of the music domain, such as dependency on contextual factors, e.g., time or device user interacted, and rapid dynamics of user's interest. A few recent studies have aimed to alleviate such difficulties in the music domain. Hansen et al. \cite{cosernn} reflected past consumption and contextual factors (e.g., the time of the day, the device used to access the service, and stream sources). Moreover, Fazelnia et al. \cite{fsvae} recently proposed utilizing user representations that consider long-term, stable interests and rapidly shifting current preferences. Despite their progress in providing more personalized experiences, they overlook the prominent and essential characteristic that uniquely appears in the music domain: the \textit{shuffle play} environments, where a set of tracks within a session are randomly provided. 

However, shuffle play sessions should not be overlooked as they take a substantial proportion (i.e., 40.2\%) of the total sessions (See Figure ~\ref{fig:degrade_shuffle} (a)), implying that the shuffle play service is highly preferred by a large number of users and is frequently utilized in real-world scenarios.
Moreover, an appropriate recommendation of a new track in shuffle play sessions would mitigate listening monotony and present serendipity in the user's auditory journey \cite{serendipityshuffle}. This perspective is corroborated by Spotify's ``Smart Shuffle\footnote{\url{https://support.spotify.com/us/article/shuffle-play/}},'' which is a recently launched service that aims to provide personalized shuffled sessions.

This work provides accurate recommendations for shuffle play sessions in the music domain. 
To begin with, we validate the effectiveness of existing state-of-the-art SBR models in each test environment. More precisely, we train SBR models (i.e., SRGNN \cite{srgnn}, FMLP \cite{fmlp}, and CL4SRec \cite{cl4srec}) and a recent music recommender system (i.e., CoSeRNN \cite{cosernn}) on Music Streaming Sessions Dataset (MSSD)~\cite{mssd} provided by Spotify. We then evaluate their performance in detail in each test environment (as depicted in Figure~\ref{fig:degrade_shuffle} (b)). Notably, although the recommender models encountered numerous shuffle play sessions during training, they all performed poorly in predicting the next track in such sessions. Providing a satisfying recommendation in shuffle play sessions is extremely challenging compared to non-shuffle play sessions.

\begin{figure}[!t]
    \includegraphics[width=1\columnwidth]{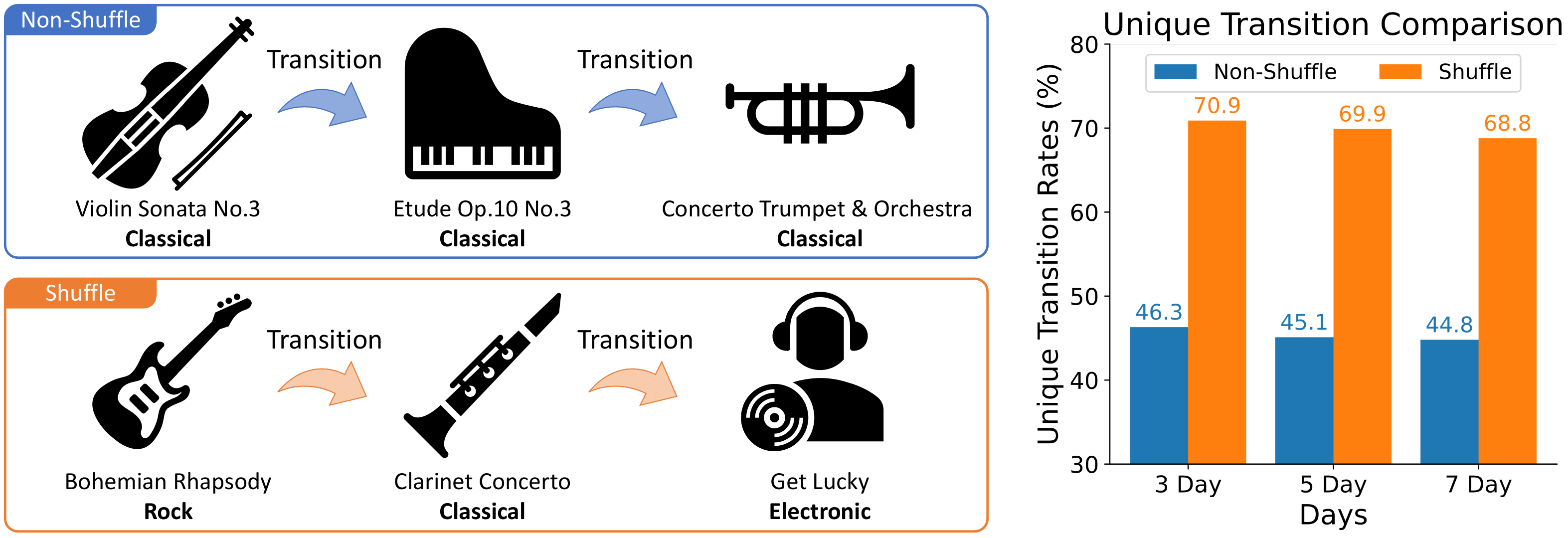}
    \vspace{-5ex}
    \caption[Unique transition rates]{\label{fig:unique_transition} Comparison of unique transition rates between non-shuffle and shuffle play sessions in MSSD dataset. Unique transition rates $=\frac{\text{\# of Unique Transitions}}{\text{\# of Total Transitions}}(\%)$}
    \vspace{-4ex}
\end{figure}

Then, an important question arises: Why do {shuffle play} sessions act as a bottleneck in building effective music recommender systems? The main clue lies in the difference in the music transition patterns between non-shuffle and shuffle play sessions. More precisely, as non-shuffle play sessions are rooted in a user's sequential history that reflects the user's taste, transitions within these sessions are unlikely to undergo dramatic shifts. For instance, if a user prefers classical music, the transitions within the session would be around similar classical music tracks. As a result, the number of \textit{unique transitions}\footnote{A unique transition indicates a transition between tracks that appears only once.} would be small. However, in shuffle play sessions, where the next music is randomly provided to users, the number of unique transitions would be large compared to those of the non-shuffle case, as illustrated in Figure ~\ref{fig:unique_transition}. 
Specifically, the unique transition rate in shuffle play sessions is considerably higher—about 1.5 times—than that of non-shuffle play sessions. Hence, we argue that such a unique transition poses a significant challenge for training SBR models, as the models need to accommodate rare and previously unseen transition patterns during training.

To this end, we propose a novel framework for training SBR, named \textbf{Mu}sic Recommender System with \textbf{S}huffle Play Recommendation \textbf{E}nhancement (\proposed), specifically designed to tackle the inherent challenges posed by shuffle play sessions in the music recommendation.~\proposed~captures the potential sequential information from shuffle play sessions using a novel session augmentation method, called transition-based augmentation.
The main idea is to insert more frequently appearing transitions to reduce a considerable proportion of unique transitions, which results in more effective use of shuffle play sessions.
Moreover, to obtain a robust unified encoder that works within diverse environments, we employ another augmentation method called reorder-based augmentation for non-shuffle play sessions, whose main idea is to mimic the shuffle-play environment.

After applying augmentations on shuffle and non-shuffle play sessions, we employ a self-supervised learning framework to maximize the agreement between the original and the augmented sessions. 
To further facilitate the alignment of representations between the two views, we introduce two fine-grained matching strategies, i.e., the item-based matching strategy that allows the identical items between the two views to be close in the embedding space, and the similarity-based matching strategy that supplements the alignment of similar embeddings between the views based on the nearest neighbors of each track.

Through extensive experiments, we demonstrate that~\proposed~outperforms recent SBR models and existing music recommender systems in predicting the next track, evaluated under various settings. To the best of our knowledge, this is the first work that attempts to enhance prevailing shuffle-play environments in the music domain in terms of training and inference.\newline

In summary, our contributions are three-fold:
\renewcommand\labelitemi{\tiny$\bullet$}
\begin{itemize}[leftmargin=3mm]
    \item {We study the characteristic of the shuffle play sessions in the music recommender system and find that a large portion of unique transitions within shuffle play sessions poses a significant challenge for training existing SBR models.} To this end, we propose a novel session augmentation method, called transition-based augmentation, that reduces the proportion of unique transitions of the shuffle play sessions.
    \item Our proposed method,~\proposed, employs self-supervised learning to maximize the agreement between the original and augmented sessions.
    To further facilitate the alignment of the representations between the two views, we devise two fine-grained matching strategies, i.e., item- and similarity-based matching strategies.
    \item Through extensive experiments, we demonstrate the superiority of~\proposed~over recent session-based recommender models and a music recommender model in the next track prediction task in a real-world music streaming dataset, MSSD.
\end{itemize}

\begin{figure*}[!t]
    \includegraphics[width=1.5\columnwidth]{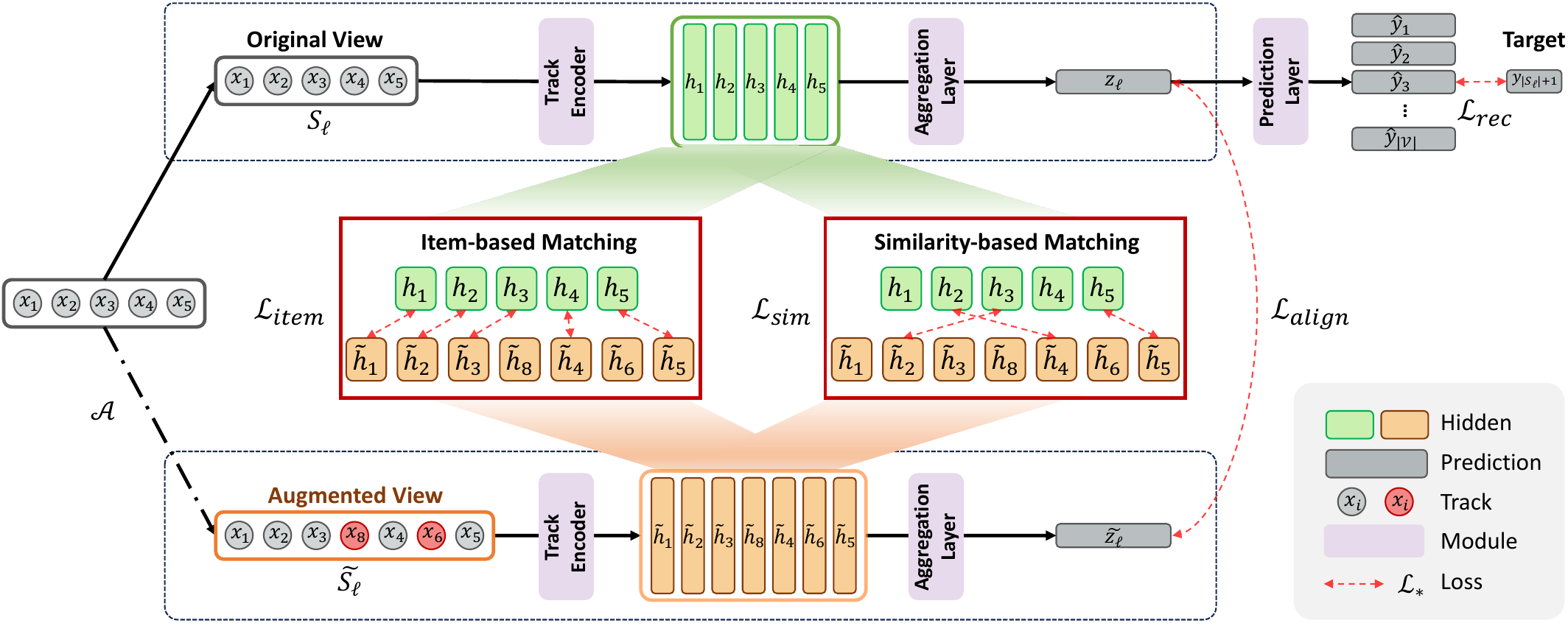}
    \vspace{-2ex}
    \caption[Overall architecture of~\proposed]{\label{fig:main_figure} Overall architecture of~\proposed. {Given a session, we first generate an augmented view via transition-based augmentation in order to alleviate the unique transition problem. 
    Subsequently, we employ item- and similarity-based matching to obtain a robust and unified encoder that can handle both shuffle and non-shuffle play sessions.} 
    }
    \vspace{-3ex}
\end{figure*}

\section{Related Work}
\noindent{\textbf{Session-based Recommendation (SBR).}}
Recurrent Neural Networks (RNNs) have found utility in session-based recommendation (SBR), leveraging their capability to model sequential data. For instance, Hidasi et al. \cite{gru4rec, gru4rec1} adapted the Gated Recurrent Unit (GRU) with the ranking loss function, aiming to predict the subsequent item of a session in SBR. NARM \cite{narm} extended GRU4Rec by incorporating an attention mechanism to capture the user's main purpose in the current session. 
Inspired by promising results in the Natural Language Processing (NLP) domain \cite{transformer}, some SBR models have embraced the self-attention mechanism. For example, SASRec \cite{sasrec} adopted the self-attention mechanism to capture both local and global interests. 
Recently, Graph Neural Networks (GNNs) have been proposed to derive item embedding for SBR using their ability to encode the relation between nodes. In particular, SR-GNN \cite{srgnn} leveraged gated GNN \cite{ggnn} to update item embeddings, considering complex item transitions, and it employed the attention mechanism akin to NARM's. 
Meanwhile, a few SBR models have focused on specific problems of implicit feedback, such as the noise of the sequential data \cite{fmlp}. FMLP \cite{fmlp} incorporated filtering algorithms from signal processing to minimize the noise in a session.
Despite these advancements, none of these approaches adequately address the specific challenges associated with the music domain, such as the existence of shuffle play sessions. In contrast, our proposed model is designed to handle shuffle play sessions using item-matching and similarity-matching modules with a self-supervised learning framework. To our knowledge, this work is the first to explicitly address the challenges associated with shuffle play sessions.

\noindent{\textbf{{Music Recommendation.}} In the music domain, the primary objective of the recommender system is to enrich the user experience by suggesting relevant tracks or artists that align with users' preferences. 
While achieving this goal, the discrepancy between industrial applications and academic research has been magnified due to the industry’s exclusive access to online streaming data via their platforms.
To close the gap, Spotify has partially released the Music Streaming Sessions Dataset (MSSD) \cite{mssd} and even hosted a sequential skip prediction challenge\footnote{\url{https://www.aicrowd.com/challenges/spotify-sequential-skip-prediction-challenge}}. It has led to numerous studies \cite{skipbehavior, skip1, skip2, skip3, skip4, skip5} aiming to make better use of implicit user feedback, i.e., skips, to enhance user experience. However, these studies primarily concentrate on the skip prediction task. This task is a binary classification that operates under the assumption of having a set of items users are certain to consume in the near future, rendering it a relatively simple task. Although a few research \cite{cosernn, fsvae} has attempted to precisely predict the subsequent item a user will interact with, its broad impact is rather limited due to restricted access to the comprehensive dataset that includes user demographics or exact timestamps. Moreover, they overlook the shuffle play environments, which frequently co-occur with non-shuffle plays but negatively impact the overall training of the recommender system due to the inherent randomness involved. In this regard, we propose a novel framework for training SBR for the music domain that predicts the next track while considering shuffle play environments, arguably a complex and challenging task.

\noindent{\textbf{Self-supervised Learning (SSL).}} SSL has recently shown remarkable performance across various domains, including Computer Vision (CV) \cite{simclr, moco, simsiam, byol, barlow, vicreg, vicregl}, Natural Language Processing (NLP) \cite{cert, simcse, clear}, and recommender systems \cite{s3rec, cl4srec, duorec}. SSL is a representation learning method that leverages supervision signals intrinsically generated from the data, eliminating the dependency on human-provided labels.
Specifically, CL4SRec \cite{cl4srec} proposed three data-level augmentations for the item sequence data and applied contrastive learning to enhance the user representation. Furthermore, DuoRec \cite{duorec} suggested a model-level augmentation based on dropout \cite{dropout} and applied supervised contrastive loss \cite{supcon} as a regularizer to alleviate the representation degeneration problem.
Unlike these approaches, mainly aiming at music recommender systems, our work proposes a novel data-level augmentation method that reflects the nature of the music domain.

\section{Proposed Framework: ~\proposed}
In this section, we first formulate the problem of session-based recommendation (SBR) and self-supervised learning framework in Section \ref{sec:3.1}. 
Then, we describe the architecture of~\proposed.
Specifically, in Section \ref{sec:3.2}, we propose a novel session augmentation method for enhancing the robustness of the session encoder to shuffle-play sessions.
In Section \ref{sec:3.3}, we introduce fine-grained matching strategies between the original and augmented sessions, followed by the description of the aggregation and prediction layer in Section \ref{sec:3.4}.
Finally, we summarize the overall training process in Section \ref{sec:3.5}.

\subsection{Preliminaries}
\label{sec:3.1}
\subsubsection{\textbf{Problem Statement}}
The objective of SBR is to predict a user's future interactions, specifically the subsequent track (i.e., item). Given a session index $\ell$ with $N$ sessions in total, a session $S_{\ell}=[x_1, x_2, ..., x_{|S_{\ell}|}]$ is composed of a sequence of tracks, where $x_t \in \mathcal{V}$ is the $t$-th track in the session and $\mathcal{V}$ is the set of all tracks in the data. 
The goal of is to predict the next track (i.e., $x_{|S_{\ell}|+1}$) given the past interactions $[x_1, \dots, x_{|S_{\ell}|}]$ in a given session $S_{\ell}$. We aim to recommend top-\textit{K} tracks for each session, given that user identity information is inaccessible due to the inherent nature of anonymous sessions.

\subsubsection{\textbf{Session-based Recommendation}}
\noindent Given an input session $S_\ell=[x_1, x_2, ..., x_{|S_{\ell}|}]$, recommender systems generally embed the tracks into embedding vectors, $\mathbf{E}_\ell=[\mathbf{e}_1, \mathbf{e}_2, ..., \mathbf{e}_{|S_{\ell}|}]$, where $\mathbf{e}_t \in \mathbb{R}^d$ is the $d$-dimensional embedding of the $t$-th track.
Then, a track encoder $f$ produces the representation of each track, $\mathbf{H}_{\ell}=f(\mathbf{E}_\ell)=[\mathbf{h}_1, \mathbf{h}_2, ..., \mathbf{h}_{|S_{\ell}|}]$, where $\mathbf{h}_t \in \mathbb{R}^d$, by modeling the interaction among tracks. Then, an aggregation layer $g$ aggregates the track representations into a session representation $\mathbf{z}_\ell=g(\mathbf{H}_\ell)$ where $\mathbf{z}_\ell \in \mathbb{R}^d$.
Given the session representation $\mathbf{z}$, a prediction layer with softmax operation produces the prediction probability for all tracks, $\hat{\mathbf{y}} = \{ \hat{y}_1, \hat{y}_2, ..., \hat{y}_{|\mathcal{V}|}\}$. The training loss $\mathcal{L}_{rec}$ can be a classification loss such as cross-entropy loss. Lastly, the model recommends top-$K$ tracks based on the prediction probability $\hat{\mathbf{y}}$. 

\subsubsection{\textbf{Self-Supervised Learning (SSL) Framework}}
\noindent To leverage shuffle-play sessions during training, we propose a SSL framework, as shown in Figure \ref{fig:main_figure}.
The framework takes a session $S_{\ell}$ as input, which can be either a shuffle or non-shuffle play session. Given the input session $S_{\ell}$, an augmentation operation $\mathcal{A}$ augments the input session based on the transition frequency. As a result, the recommender system better captures users' preferences from the shuffle play sessions to provide more accurate recommendations.
More formally, we embed the tracks into embedding vectors, $\mathbf{E}_{\ell}$ and $\tilde{\mathbf{E}}_{\ell}$, from the original and augmented sessions (i.e., $S_{\ell}$ and $\tilde{S}_{\ell}$), respectively.
Then, a track encoder $f$ produces track representations by modeling the interaction among the tracks in each session, i.e., $\mathbf{H}_{\ell} = f(\mathbf{E}_{\ell})$ and $\tilde{\mathbf{H}}_{\ell} = f(\tilde{\mathbf{E}}_{\ell})$. 
The aggregation layer $g$ aggregates the track representations into session representations, i.e., $\mathbf{z}_\ell = g(\mathbf{H}_{\ell})$ and $\tilde{\mathbf{z}}_\ell = g(\tilde{\mathbf{H}}_{\ell})$. A basic SSL approach aligns the final representations (i.e.. $\mathbf{z}_\ell$ and $\tilde{\mathbf{z}}_\ell$) by increasing their similarity, resulting in the alignment loss (i.e., $\mathcal{L}_{align}$).
We note that we employ a shared track encoder $f$ and a shared aggregation layer $g$ in both branches.

\subsection{\textbf{Transition-based Augmentation}}
\label{sec:3.2}
Transition-based augmentation aims to enrich the sequential information in a given shuffle play session. To this end, we consider the transition frequency between items from all the sessions as an essential criterion for distinguishing shuffle and non-shuffle play sessions, as shown in Figure \ref{fig:augmentation}. 
We first demonstrate how we obtain a transition matrix and propose a novel session augmentation method conducted based on the transition matrix.

\begin{figure}[!t]
    \centering
    \includegraphics[width=0.9\columnwidth]{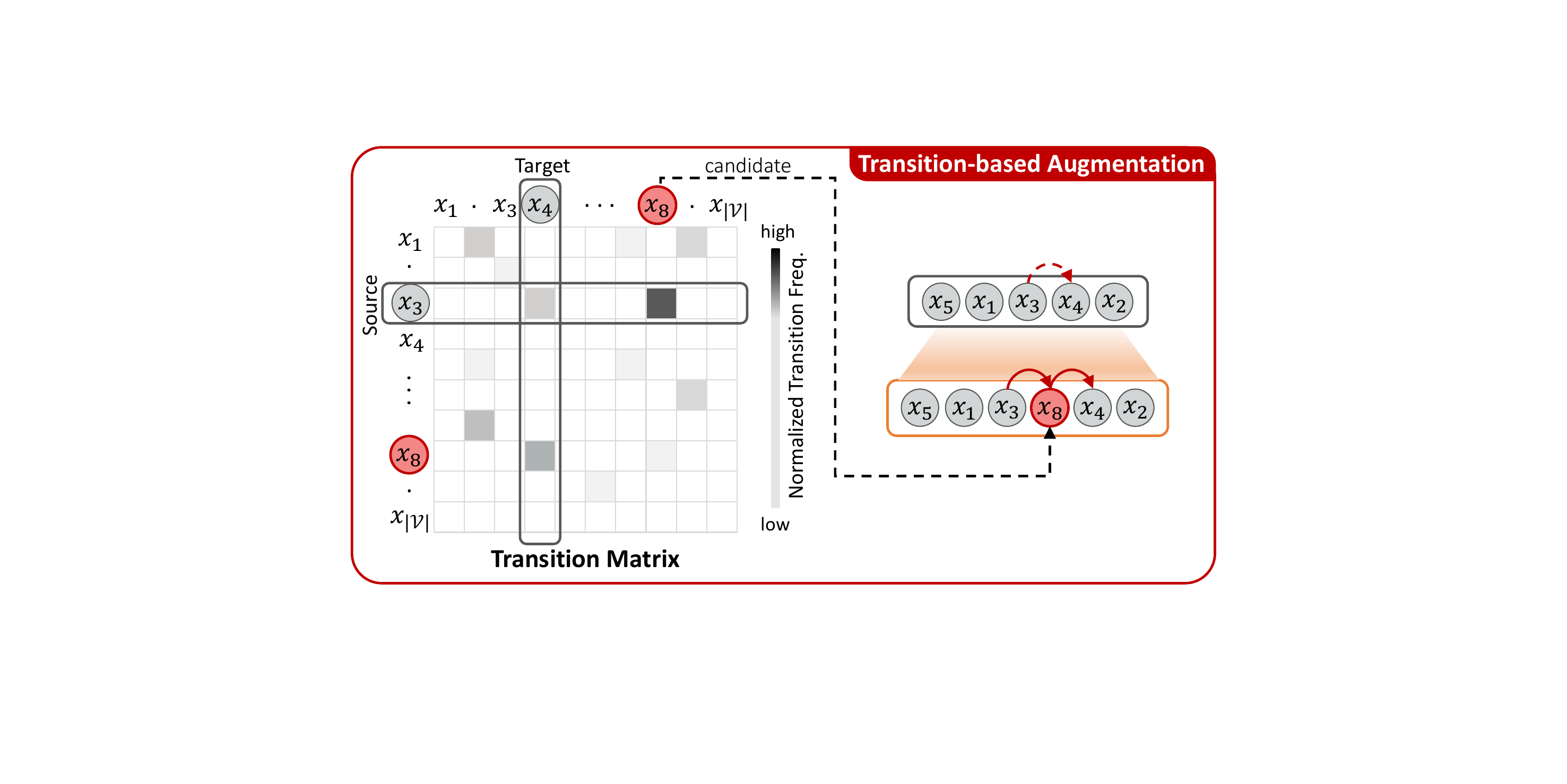}
    \vspace{-3ex}
    \caption[Transition-based augmentation]{\label{fig:augmentation} Our proposed transition-based augmentation showing an example of inserting a track $x_8$ between $x_3$ and $x_4$.}
    \vspace{-3ex}
\end{figure}

\noindent{\textbf{Transition Matrix.}} As shown in Figure~\ref{fig:unique_transition}, the main challenge inherent in the shuffle play sessions is their excessive amount of unique transitions within a session. 
To address the problem of excessive unique transitions, we introduce non-unique transition patterns observed across all sessions to shuffle play sessions. By doing so, we effectively mitigate the unique transition patterns inherent in shuffle play sessions, thereby unlocking the potential for leveraging these sessions during the training process.
More formally, we first generate a transition frequency matrix $\mathbf{T} \in \mathbb{R}^{|\mathcal{V}| \times |\mathcal{V}|}$, by collecting all transitions observed in the entire sessions as follows:
\begin{equation}
    \mathbf{T}_{i,j} = \sum_{\ell=1}^{N} \sum_{t=1}^{|S_{\ell}|-1} \mathbbm{1}([x_t, x_{t+1}]=[x_i, x_j]), \quad \forall i,j \leq |\mathcal{V}|
\end{equation}
where $\mathbf{T}_{i,j}$ denotes the frequency of transition from source track $x_i$ to target track $x_j$,  $\mathbbm{1}(a=b)$ denotes indicator function which outputs 1 if $a=b$ else 0, and $N$ is the total number of sessions.
We also take the logarithm to each value in $\mathbf{T}$ as the transition frequency of certain pairs, e.g., the transition between popular tracks, tends to be much higher than that of the remaining cases\footnote{Here, we ensure log transformation is applied to non-zero frequency values.}, which may incur the long-tail problem \cite{log1, log2, log3, melt, lte4g}.
We then normalize the log-transformed matrix from the following two perspectives:
\begin{equation}
\begin{gathered}
    \bar{\mathbf{T}}_{i,\boldsymbol{\cdot}} = \frac{{\mathbf{T}}_{i,\boldsymbol{\cdot}}}{\sum_{j=1}^{|\mathcal{V}|}\mathbf{T}_{i,j}}, \;\; \forall i \leq |\mathcal{V}|, \quad
    \bar{\mathbf{T}}_{\boldsymbol{\cdot}, j} = \frac{{\mathbf{T}}_{\boldsymbol{\cdot}, j}}{\sum_{i=1}^{|\mathcal{V}|}\mathbf{T}_{i,j}}, \;\; \forall j \leq |\mathcal{V}|
\end{gathered}
\end{equation}
where $\bar{\mathbf{T}}_{i,\boldsymbol{\cdot}}$ and $\bar{\mathbf{T}}_{\boldsymbol{\cdot},j}$ denote the row-wise (i.e., source-wise) and column-wise (i.e., target-wise) normalized transition matrices, respectively. This results in the Markov Chain Transition Matrices \cite{markov}, where the transition probability of each source and target node sums to one. This normalization enables us to interpret the transition matrix in terms of the probability distribution matrix and take a stochastic approach while augmenting a given session.

\noindent{\textbf{Transition-based Insertion.}} We now propose a novel session augmentation method to handle shuffle play sessions for music recommendation. The main idea is to insert frequently appearing transitions that could exist in a session. The primary goals of the augmentation are: (1) to reduce the excessive amount of unique transitions in shuffle play sessions and (2) to expose the session encoder to more diverse environments, thereby better accommodating shuffle play sessions. 
More precisely, our proposed augmentation method determines which items to be inserted at which locations in a given session. Here, the key idea lies in not inserting any random items but inserting relevant items that are likely to appear considering its back-and-forth context, i.e., source and target. For a clear and comprehensive understanding, the reader is encouraged to refer to Figure~\ref{fig:augmentation}, which illustrates a toy example of inserting $x_8$ between $x_3$ and $x_4$. Specifically, as $x_3$ has a high transition probability to $x_8$, and $x_8$ has a high transition probability to $x_4$, we insert $x_8$ between $x_3$ and $x_4$.

For the efficiency of computation, we formally describe the insertion process of multiple tracks. Given an input session $S_{\ell}=[x_1,x_2,\dots,x_{|S_{\ell}|}]$, we have $|S_{\ell}|-1$ candidate slots between the tracks for insertion. 
Thus, we set source tracks appearing before insertion (i.e., $S_{\ell}^{\textsf{s}}=[x_1,x_2,\dots,x_{|S_{\ell}|-1}]$) and target tracks appearing after insertion (i.e.,  $S_{\ell}^{\textsf{t}}=[x_2,x_3,\dots, x_{|S_{\ell}|}]$). 
Then, we obtain the transition matrices for source and target tracks such that:
\begin{equation}
    \bar{\mathbf{T}}_{S_{\ell}^{\textsf{s}},\boldsymbol{\cdot}} \in \mathbb{R}^{(|S_{\ell}|-1)\times|\mathcal{V}|}, \quad \bar{\mathbf{T}}_{\boldsymbol{\cdot},S_{\ell}^{\textsf{t}}} \in \mathbb{R}^{|\mathcal{V}|\times (|S_{\ell}|-1)}.
    \label{eqn:transprob}
\end{equation}
Figure \ref{fig:augmentation} shows an example of insertion between $x_3$ and $x_4$, while Eq. \ref{eqn:transprob} considers all cases of insertion. We then obtain the probability of candidate tracks for insertion between source and target tracks as follows:
\begin{equation}
    \mathbf{P}_{S_{\ell},\boldsymbol{\cdot}} = \bar{\mathbf{T}}_{S_{\ell}^{\textsf{s}},\boldsymbol{\cdot}} \odot {\bar{\mathbf{T}}_{\boldsymbol{\cdot}, S_{\ell}^{\textsf{t}}}}^{\top}
\end{equation}
where $\mathbf{P}_{S_{\ell},\boldsymbol{\cdot}} \in \mathbb{R}^{(|S_{\ell}|-1)\times|\mathcal{V}|}$ denotes a matrix of potential candidates that could be inserted between tracks in a given session, with values obtained via Hadamard product, $\odot$, of two subsets of transition probability matrices\footnote{For the implementation, transition matrix is stored as sparse tensors regarding its high sparsity, hence the memory cost is notably low.}. It is important to note that as both matrices consist of Markov Chain Transition probabilities, the potential candidates would contain a value that naturally considers its stochastic nature, conditioned on both the source track and the target track. We also apply row-wise softmax to ensure the probability distribution, i.e., $\bar{\mathbf{P}}_{S_{\ell},\boldsymbol{\cdot}} = \textsf{softmax} (\bar{\mathbf{T}}_{S_{\ell}^{\textsf{s}},\boldsymbol{\cdot}} \odot {\bar{\mathbf{T}}_{\boldsymbol{\cdot}, S_{\ell}^{\textsf{t}}}}^{\top})$.

Based on the potential candidates obtained from transition matrices, we  sample a candidate track to be inserted in each interval of the sequence as follows:
\begin{equation}
\label{eqn:mult}
\begin{gathered}
    \mathbf{c}_{i} = 
    \begin{cases}
        \textsf{Multinomial}({{\bar{\mathbf{P}}}_{S_{\ell}}[i,:]}), & \text{if} \; \textsf{sum}({{\bar{\mathbf{P}}}_{S_{\ell}}[i,:]}) > 0 \\
        \varnothing, & \text{otherwise} 
    \end{cases}
    , \forall i \leq |S_{\ell}|-1
\end{gathered}
\end{equation}
where $\mathbf{c}_i$ is the $i$-th element of $\mathbf{c} \in \mathbb{R}^{|S_{\ell}|-1}$, which is initialized as zeros then replaced with a sample obtained from Multinomial Distribution with event probabilities, ${{\bar{\mathbf{P}}}_{S_{\ell}}[i,:]}$, given at least one candidate, i.e., a potential track that is associated to both a transition from the source track and a transition to the target track exists ($\textsf{sum}({{\bar{\mathbf{P}}}_{S_{\ell}}[i,:]}) > 0$). An example of such a potential track is $x_8$ in Figure~\ref{fig:augmentation}. Finally, we obtain the augmented session $\tilde{S}_{\ell}$ by inserting the sampled tracks $\mathbf{c}$ into the original session $S_{\ell}$. {Here, when new tracks are inserted between each track in the original session, we employ the augmented session unless its length surpasses the maximum session length. However, if the number of candidate tracks exceeds the available slots (i.e., $X - (|S_{\ell}|-1)$, where $X$ is the maximum session length), we randomly pick tracks from the candidates to ensure that each track has an equal opportunity for integration into the session.} In summary, the augmented session alleviates the unique transition problem by inserting relevant transitions.

\noindent\textbf{Discussions on non-shuffle play sessions.} Heretofore, we mainly discussed augmenting shuffle play sessions. However, we can also benefit from applying augmentations to the non-shuffle play sessions, as shown by an existing work~\cite{cl4srec}. Here, we opted not to apply transition-based augmentation to non-shuffle play sessions, given that their transition patterns are not as unique as those of shuffle play sessions. Furthermore, we empirically observed that using transition-based augmentation for non-shuffle play sessions did not result in a performance gain. 
Instead, we apply reorder-based augmentation for non-shuffle play sessions, which randomly reorders the tracks within a session, thereby mimicking the shuffle play environment.
By exposing the reordered non-shuffle play sessions to the session encoder, we obtain a robust and unified encoder invariant to the shuffles.

\subsection{\text{Item- and Similarity-based Matching}}
\label{sec:3.3}
In this section, we propose fine-grained matching strategies, i.e., item- and similarity-based matching, to better align the original and augmented session. 
As illustrated in Figure~\ref{fig:main_figure}, we obtain an augmented session $\tilde{S}_{\ell}$ of the input session $S_{\ell}$ through an augmentation operation $\mathcal{A}$.
After looking up the embedding vectors $\mathbf{E}_{\ell} \in \mathbb{R}^{|S_{\ell}| \times d}$,
and $\tilde{\mathbf{E}}_{\ell} \in \mathbb{R}^{|\tilde{S}_{\ell}| \times d}$
for each item within each view, the track encoder $f$ generates track representations, i.e., $\mathbf{H}_{\ell} \in \mathbb{R}^{|{S}_{\ell}| \times d}$ and $\tilde{\mathbf{H}}_{\ell} \in \mathbb{R}^{|\tilde{S}_{\ell}| \times d}$, corresponding to each view.
We now delineate the matching strategies.

\subsubsection{{\textbf{Item-based Matching}}}
The augmentations make the encoder generate different hidden representations of the same items due to the differing adjacent items. Nonetheless, we aim to make the encoder to be invariant to such augmentations. The item-based matching ensures the alignment between the two views' hidden representations derived from the same items. 
Let the items from original session $S_{\ell}$ be $\mathbf{I}_\ell = \{ x_i | x_i \in S_{\ell}\}$ and the items from augmented session $\tilde{S}_{\ell}$ be $\tilde{\mathbf{I}}_\ell = \{ x_i | x_i \in \tilde{S}_{\ell}\}$. The item-based matching loss function, Mean Squared Error, is defined as follows:
\begin{equation}
    \mathcal{L}_{item} = \frac{1}{|\mathbf{I}_\ell|} \sum_{x_t \in \mathbf{I}_\ell}\sum_{x_k \in \tilde{\mathbf{I}}_\ell}{\mathbbm{1}(x_t = x_k)\|\mathbf{h}_{t} - \mathbf{\tilde{h}}_{k}\|^2},
\end{equation}
where $\mathbbm{1}(a=b)$ is the indicator that produces 1 if $a=b$ and 0 otherwise, $\mathbf{h}_{t}, \mathbf{\tilde{h}}_{k} \in \mathbb{R}^{d}$ are representations of $t$-th track in the original session and the $k$-th track in the augmented session, respectively.

\subsubsection{{\textbf{Similarity-based Matching}}}
In addition to the item-based matching strategy, inspired by the importance of neighborhood information in recommender systems~\cite{bell2007improved}, we employ similarity-based matching that aligns representations of similar items. 
Unlike item-based matching, which focuses on aligning the representations of the same item from the two views, 
similarity-based matching determines the nearest neighbor of each item in one view from the items in the other view. To accomplish this, we first calculate the Euclidean distance in the embedding space between all pairs of tracks in the original session $S_{\ell}$ and the augmented session $\tilde{S}_{\ell}$, then select the nearest neighbor (\textsf{NN}) track for each track representation as follows:
\begin{equation}
\mathcal{P}(\mathbf{H}_{\ell}, \tilde{\mathbf{H}}_{\ell}) = \{\, (\mathbf{h}_i, \textsf{NN}(\mathbf{h}_{i}, \tilde{\mathbf{H}}_{\ell})) \, | \, \mathbf{h}_i \in \mathbf{H}_{\ell} \}
\end{equation}
where $\mathcal{P}(\mathbf{H}_{\ell}, \tilde{\mathbf{H}}_{\ell})$ is the set of track pairs in which one is from the original session and the other is its nearest neighbor from the augmented session. $|\mathcal{P}(\mathbf{H}_{\ell}, \tilde{\mathbf{H}}_{\ell})| = |{S}_{\ell}|$, and $\mathsf{NN}(\mathbf{h}_{i}, \tilde{\mathbf{H}}_{\ell})$ returns the representation of the nearest neighbor track of the $i$-th track of the original session (i.e., $\mathbf{h}_{i}$) among all tracks in the augmented session (i.e., $\tilde{\mathbf{H}}_{\ell}$). 
Then, we select the top-$\kappa$ tracks with the lowest distance denoted as $\mathcal{P}^\kappa(\mathbf{H}_{\ell}, \tilde{\mathbf{H}}_{\ell})$, where $|\mathcal{P}^\kappa(\mathbf{H}_{\ell}, \tilde{\mathbf{H}}_{\ell})|=\kappa$ and $\mathcal{P}^\kappa(\mathbf{H}_{\ell}, \tilde{\mathbf{H}}_{\ell}) \subset \mathcal{P}(\mathbf{H}_{\ell}, \tilde{\mathbf{H}}_{\ell})$. The purpose of introducing top-$\kappa$ selection is to ensure that only similar pairs are considered so that this process complements the item-based matching.
Likewise, we consider the top-$\kappa$ nearest neighbors in the perspective of the augmented session, i.e., $\mathcal{P}^\kappa(\tilde{\mathbf{H}}_{\ell}, \mathbf{H}_{\ell})$.
The loss function for similarity-based matching is defined as follows:
\begin{multline}
     \mathcal{L}_{sim} = \sum_{(\mathbf{h}_i, \textsf{NN}(\mathbf{h}_{i}, \tilde{\mathbf{H}}_{\ell})) \in \mathcal{P}^\kappa} \|\mathbf{h}_i - \textsf{NN}(\mathbf{h}_{i}, \tilde{\mathbf{H}}_{\ell})  \|^2 + \\
     \sum_{(\tilde{\mathbf{h}}_i, \textsf{NN}(\tilde{\mathbf{h}}_{i}, {\mathbf{H}}_{\ell})) \in \tilde{\mathcal{P}}^\kappa} \|\tilde{\mathbf{h}}_i - \textsf{NN}(\tilde{\mathbf{h}}_{i}, {\mathbf{H}}_{\ell})  \|^2
\end{multline}
where $\mathcal{P}^\kappa=\mathcal{P}^\kappa(\mathbf{H}_{\ell}, \tilde{\mathbf{H}}_{\ell})$ and $\tilde{\mathcal{P}}^\kappa=\tilde{\mathcal{P}}^\kappa(\tilde{\mathbf{H}}_{\ell}, {\mathbf{H}}_{\ell})$ for simplicity.
It is worth noting that incorporating similarity-based matching from the beginning may interfere with the training, as the representations are not yet established.
Therefore, after some warm-up epochs with only the item-based matching, we start the similarity-based matching, aiming to obtain meaningful representations\footnote{In experiments, we used the first epoch of the training as the warm-up epoch.}. 

\subsubsection{\textbf{Regularization.}} 
To avoid the representation collapse problem prevalent in the self-supervised learning framework, we employ the regularization strategy introduced in VICReg \cite{vicreg}: $\mathcal{L}_{VICReg} = \lambda \cdot s(\mathbf{H}_{\ell}, \tilde{\mathbf{H}}_{\ell}) + \mu [v(\mathbf{H}_{\ell}) + v(\tilde{\mathbf{H}}_{\ell})] + \nu [c(\mathbf{H}_{\ell}) + c(\tilde{\mathbf{H}}_{\ell})]$, where $s$, $v$, and $c$ are the invariance, variance, and covariance terms, respectively, and $\lambda$, $\mu$, and $\nu$ are scalar coefficients terms. Therefore, the final loss function of these matching is defined as follows:
\begin{equation}\label{loss:matching}
    \mathcal{L}_{matching} = \mathcal{L}_{item} + \mathcal{L}_{sim} + \mathcal{L}_{VICReg}.
\end{equation}

\subsection{Aggregation and Prediction Layer}
\label{sec:3.4}
\sloppy After the track encoder models all interactions among tracks in a session, we leverage the track representations $\mathbf{H}_{\ell} = [\mathbf{h}_1, \mathbf{h}_2, \dots, \mathbf{h}_{|S_{\ell}|}]$ to aggregate both long-term preference and current interests of the session. 
In the aggregation layer $g$, We first consider the last track representation $\mathbf{h}_{|S_{\ell}|}$ as the local embedding $\mathbf{z}_{\ell}^{(\text{local})}$ of the session, i.e., $\mathbf{z}_{\ell}^{(\text{local})} = \mathbf{h}_{|S_{\ell}|}$. 
Then, we derive the global embedding $\mathbf{z}_{\ell}^{(\text{global})}$ from all track representations. Here we adopt Bahdanau attention \cite{bahdanauattention} by following the previous works \cite{narm, stamp, srgnn}:
\begin{equation}
    \mathbf{z}_{\ell}^{(\text{global})} = \sum_{i}^{|S_{\ell}|} {\beta_{i} \mathbf{h}_{i}}, \,\,\beta_{i} = \mathbf{W}_{1}^{T}\sigma(\mathbf{W}_{2} \mathbf{h}_{i} + \mathbf{W}_{3} \mathbf{h}_{|S_{\ell}|} + \mathbf{b}) 
\end{equation}
where learnable parameters $\mathbf{W}_{1} \in \mathbb{R}^{d}$, $\mathbf{W}_{2}, \mathbf{W}_{3} \in \mathbb{R}^{d \times d}$ and bias $\mathbf{b}$ control the weight of track representations, and $\sigma$ is an activation function.
Finally, we concatenate and transform local $\mathbf{z}_{\ell}^{(\text{local})}$ and global $\mathbf{z}_{\ell}^{(\text{global})}$ embedding to a $d$-dimensional embedding: $\mathbf{z}_\ell = \mathbf{W}_{4}(\mathbf{z}_{\ell}^{(\text{local})} \oplus \mathbf{z}_{\ell}^{(\text{global})})$, where $\oplus$ is a concatenate operator and $\mathbf{W}_{4} \in \mathbb{R}^{d \times 2d}$. After feeding the augmented track representation into this aggregation layer, we obtain the augmented session representation $\tilde{\mathbf{z}}_\ell = g(\tilde{\mathbf{H}}_\ell)$. To align the two session representations, we employ the self-supervised loss introduced in VICReg \cite{vicreg}: 
\begin{equation}\label{loss:align}
    \mathcal{L}_{align} = \lambda \cdot s(\mathbf{z}_{\ell}, \tilde{\mathbf{z}}_{\ell}) + \mu [v(\mathbf{z}_{\ell}) + v(\tilde{\mathbf{z}}_{\ell})] + \nu [c(\mathbf{z}_{\ell}) + c(\tilde{\mathbf{z}}_{\ell})].
\end{equation}
Given the session representation $\mathbf{z}$, a {prediction layer} computes the prediction probability $\hat{\mathbf{y}}\in\mathbb{R}^{|\mathcal{V}|}$ of the next track using softmax:
\begin{equation*}
    \hat{\mathbf{y}} = \textsf{softmax}(\mathbf{z}_{\ell}^{T}\mathbf{e}_i),
\end{equation*}
where $\mathbf{e}_i \in \mathbf{E}_{\mathcal{V}}$ is a candidate item embedding vector. For each session, we minimize the cross entropy loss defined as follows:
\begin{equation}\label{loss:rec}
    \mathcal{L}_{rec} = -\sum_{i=1}^{|\mathcal{V}|}{\mathbf{y}_i \log{(\hat{\mathbf{y}}_i)} + (1-\mathbf{y}_i) \log{(1-\hat{\mathbf{y}}_i)}},
\end{equation}
where $\mathbf{y}_i\in\mathbb{R}^{|\mathcal{V}|}$ is the one-hot vector of the target track.

\subsection{Model Training}
\label{sec:3.5}
To sum up, the final loss of~\proposed~can be expressed as follows:
\begin{equation}
    \label{loss:final}
    \mathcal{L}_{\text{final}} = \alpha \mathcal{L}_{matching} + (1-\alpha) \mathcal{L}_{align} + \mathcal{L}_{rec},
\end{equation}
\noindent where $\alpha$ is a loss-controlling hyperparameter that balances between the matching loss and alignment loss, $\mathcal{L}_{matching}$ accounts for the item- and similarity-based matching loss with a regularization loss (Eq.~\ref{loss:matching}), $\mathcal{L}_{align}$ aims for the alignment of embeddings obtained via the aggregation layer (Eq.~\ref{loss:align}), and $\mathcal{L}_{rec}$ is derived from the next track prediction task through the prediction layer (Eq.~\ref{loss:rec}).

\begin{table}[!t]
    \begin{minipage}{0.6\linewidth}{
        \small
        \caption{Statistics of datasets.}
        \vspace{-2ex}
        \resizebox{1\columnwidth}{!}{
            \begin{tabular}{lccc}
            \toprule
            \multicolumn{1}{c}{Statistics}               & MSSD-3d  & MSSD-5d  & MSSD-7d  \\ \midrule
            \# of plays               & 11,858,262 & 16,701,958 & 19,366,448 \\
            \# of shuffle play sessions    & 301,814   & 422,221   & 501,875   \\
            \# of non-shuffle play sessions & 442,726   & 618,701  & 713,300  \\
            \# of training sessions   & 613,308   & 909,818   & 1,061,274  \\
            \# of test sessions       & 131,232   & 131,104   & 153,901   \\
            \# of tracks              & 199,177   & 253,693   & 280,079   \\
            Average length           & 15.93    & 16.05    & 15.94    \\ \bottomrule
            \end{tabular}
            }
        \label{tab:statistics}
    }
    \end{minipage}
    \begin{minipage}{0.35\linewidth}{
        \centering
        \includegraphics[width=0.9\linewidth]{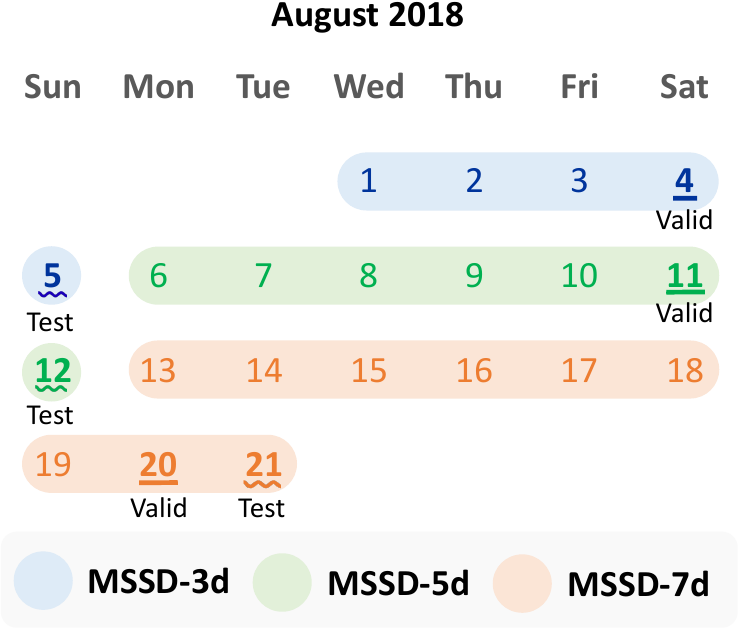}
        \vspace{-2.5ex}
        \captionof{figure}{Days split.}
        \label{fig:split}
    }
    \end{minipage}
    \vspace{-4.5ex}
\end{table}

\section{Experiments}
We first describe our experiment settings in Section 4.1 and summarize observations from the overall performance in Section 4.2. Additionally, we delineate the effectiveness of \proposed~in non-shuffle and shuffle play environments in Section 4.3. In Section 4.4, we demonstrate the efficacy of our proposed module, e.g., transition-based augmentation and fine-grained matching strategies. Finally, we show the sensitivity of each hyperparameter in Section 4.5.

\subsection{Experimental Settings}
\subsubsection{\textbf{Dataset}} We compare \proposed~with baseline methods on a large-scale real-world dataset, Music Streaming Sessions Dataset (MSSD) \cite{mssd}, from Spotify\footnote{\href{https://www.aicrowd.com/challenges/spotify-sequential-skip-prediction-challenge}{https://www.aicrowd.com/challenges/spotify-sequential-skip-prediction-challenge}}. It comprises 160 million listening sessions with 20 billion plays, accompanied by user actions. It consists of the historical logs for 66 days. 
Additionally, to ensure manageable computation time, we utilize about 50\% of the entire dataset due to its extensive size\footnote{We used data files whose file names start with log\_0, log\_1, log\_2, log\_3, and log\_4.}.
We then constitute 3 chunks of the dataset by selecting data belonging to a few days as adopted in a conventional work \cite{narm} that used chunk data of original data due to its large size. 
As illustrated in Figure \ref{fig:split}, here are the 3 chunks of the dataset:
\begin{itemize}
    \item 3 days (MSSD-3d) - training data is from 1 August 2018 to 3 August 2018, validation data is from 4 August 2018, and test data is from 5 August 2018,
    
    \item 5 days (MSSD-5d) - training data is from 6 August 2018 to 10 August 2018, validation data is from 12 August 2018, and test data is from 13 August 2018
    
    \item 7 days (MSSD-7d) - training data is from 13 August 2018 to 20 August 2018, validation data is from 21 August 2018, and test data is from 22 August 2018.
\end{itemize}

For the data preprocessing, we excluded items in the test data that do not appear in the training data, i.e., cold-start problem that is generally covered as a separate issue.
We filtered out non-premium users because they are limited to using the streaming platform as done in \cite{cosernn}.
Following the conventional works \cite{gru4rec, narm}, we also filtered out sessions containing only one track and tracks that appear less than 5 times in training data. 
\sloppy In addition, for the session $S = [x_1, x_2, \dots, x_{|S|}, x_{|S|+1}]$, we set up a series of sequences and corresponding labels $([x_1], x_2), ([x_1, x_2], x_3), \dots, ([x_1, x_2, \dots, x_{|S|}], x_{|S|+1})$, where $([\ast], \boldsymbol{\cdot})$ denotes a track sequence $[\ast]$ and next tracks $\boldsymbol{\cdot}$.
However, when generating a series of sequences and next tracks, we filtered out data instances that the user skipped the next tracks in order to recommend a track that a user will listen to.
Specifically, for a shuffle play session, we excluded all skipped tracks, even in input. Because the shuffle play session inherits the randomness, we exclude them to construct more meaningful track sequences. 
For example, given that the session is $S_{\ell}^{(Shuffle)}=[x_1, x_2, x_3, x_4, x_5]$ and the user listening behavior is [\textit{listen}, \textit{skip}, \textit{listen}, \textit{skip}, \textit{listen}], a series of sequences and corresponding labels are generated as follows: {$([x_1], x_3), ([x_1, x_3], x_5)$}.
Lastly, if the shuffle play mode (e.g., shuffle play $\rightarrow$ non-shuffle play) is changed in the middle of a session, we treat it as a shuffle play session. The detailed statistics of the dataset after the preprocessing are in Table \ref{tab:statistics}.

\subsubsection{\textbf{Evaluation Protocol}}
Since music recommender systems typically present a limited number of tracks at a time, it is important that the actual track listened to by the user is included in the top-ranked tracks of the list. Therefore, we adopt the Recall (\textbf{Recall@K}) \cite{gru4rec, narm}, Mean Reciprocal Rank (\textbf{MRR@K}) \cite{gru4rec, narm, srgnn}, and Normalized Discounted Cumulative Gain (\textbf{NDCG@K}) \cite{cl4srec, fmlp} that are frequently used when evaluating the ranking performance.

\subsubsection{\textbf{Compared Methods}}
We evaluate our proposed method compared with the following baseline methods. \textbf{General Recommender System(RS)}: 1) SimpleX \cite{simplex} is a collaborative filtering method with Cosine Contrastive Loss. We used average track embeddings as a user embedding of a session. \textbf{Classic SBR}: 2) GRU4Rec \cite{gru4rec} is RNN-based SBR (i.e., GRU) with a ranking loss function to encode the sequential information. \textbf{Attention-based SBR}: 3) NARM \cite{narm} is an RNN-based model with an attention mechanism to aggregate long- and short-term interest. 4) STAMP \cite{stamp} is an MLP-based model with a short-term attention priority module to detect shifts in user interest in a session. 5) CSRM \cite{csrm} is an extended model of NARM with outer memory to utilize collaborative signals from neighbor sessions. 6) SASRec \cite{sasrec} utilized unidirectional transformer architecture. \textbf{Graph-based SBR}: 7) SR-GNN \cite{srgnn} utilized Gated GNN to encode the item embedding and aggregate them using an attention mechanism. 8) GC-SAN \cite{gcsan} is a method for the fusion of Gated GNN and self-attention mechanism. \textbf{Self-supervised Learning-based SBR}: 9) CL4SRec \cite{cl4srec} adopted a contrastive learning framework with SASRec as a backbone. 10) DuoRec \cite{duorec} utilized model-based augmentation with supervised contrastive loss. \textbf{Recent SBR}: FMLP \cite{fmlp} replaced the self-attention module with a simple MLP with a noise-filtering algorithm from signal processing. \textbf{Music RS}: CoSeRNN \cite{cosernn} is an RNN-based (i.e., LSTM) music recommender system that utilizes context information (e.g., day, stream source). We excluded some context variables that they used due to the data availability. We utilized context (e.g., charts, personalized playlists, user collection) and item embedding as learnable embedding; the other is used as one-hot embedding. 

\subsubsection{\textbf{Implementation Details}} 
\noindent{\newline \textbf{Hyperparameters Tuning.}} We tuned the hyperparameters of the methods, including \proposed{} based on MRR@5 on the validation dataset. We then evaluated the methods with the optimal hyperparameters on the test dataset when they produced the highest MRR@5 on the validation dataset.

\noindent{\textbf{Compared Methods.}} The maximum length of sessions is 20 tracks, which is given in MSSD. We add zero padding if a session contains fewer than 20 tracks. For fair comparisons, we set the hidden dimension $d$ to 100 and batch size to 512. We also use a single layer for all graph neural networks and the transformer encoder layer. Additionally, we use a single head for the transformer encoder layer. The other model-specific hyperparameters were searched in the range reported by the authors.

\noindent{\textbf{Our Proposed Framework.}} For~\proposed, we searched the reorder probability $\gamma$ that controls the proportion of tracks in a given session in \{0.3, 0.5, 0.7, 0.9\}. For $\kappa$, which is responsible for selecting the top-$\kappa$ nearest neighbor pairs in the similarity-based matching process, we fixed it as 5. We searched  $\alpha$ in \{0.2, 0.4, 0.6, 0.8\} (Eq. \ref{loss:final}), which is the loss-controlling hyperparameter balancing the matching loss and alignment loss. For the coefficients used in VICReg regularization, we fixed $\lambda$, $\mu$, and $\nu$ to 1, 1, and 10, respectively.

\begin{table*}[t]
\caption[Overall performance comparison]{Overall performance comparison. Gen., Cls., and Rec. denote general, classic, and recent, respectively. $\Delta_b$ and $\Delta_s$ denote the relative improvement of~\proposed~over the backbone model, SRGNN, and state-of-the-art baseline, GCSAN, respectively. The performance is averaged across {5 log files in each chunk, and its standard deviation is shown in parentheses. \textbf{Bold} fonts indicate the top-ranking performance while \uline{underlining} denotes the second-place performance.
An asterisk (*) indicates the statistical significance of the improvement of our model over the top-performing baseline, as determined by a paired t-test with $p < 0.01$.}}
\vspace{-3ex}
\renewcommand{\arraystretch}{1.5}
\label{tab:performance}
\resizebox{1.95\columnwidth}{!}{
\begin{tabular}{@{}clcccccccccccccrr@{}}
\toprule
\multicolumn{2}{c}{Setting}                                                                              & Gen. RS    & Cls. SBR & \multicolumn{4}{c}{Attention-based SBR}                       & \multicolumn{2}{c}{Graph-based SBR} & \multicolumn{2}{c}{SSL-based SBR} & Rec. SBR    & Music RS      & \multicolumn{3}{c}{\textbf{Ours}}                                             \\ \cmidrule(lr){1-2} \cmidrule(lr){3-3} \cmidrule(lr){4-4} \cmidrule(lr){5-8} \cmidrule(lr){9-10} \cmidrule(lr){11-12} \cmidrule(lr){13-13} \cmidrule(lr){14-14} \cmidrule(lr){15-17}
Dataset                                                                     & \multicolumn{1}{c}{Metric} & SimpleX       & GRU4Rec         & NARM          & STAMP         & CSRM          & SASRec        & SRGNN            & GCSAN            & CL4SRec         & DuoRec          & FMLP          & CoSeRNN       & \textbf{\proposed} & \multicolumn{1}{c}{$\Delta_b$} & \multicolumn{1}{c}{$\Delta_s$} \\ \midrule
\multirow{6}{*}{\begin{tabular}[c]{@{}c@{}}MSSD\\ 3d\end{tabular}}
    & \shortstack{R@5\\\;} & \shortstack{0.1785\\\scriptsize{$(0.0019)$}} & \shortstack{0.2359\\\scriptsize{$(0.0019)$}} & \shortstack{0.3348\\\scriptsize{$(0.0021)$}} & \shortstack{0.3283\\\scriptsize{$(0.0017)$}} & \shortstack{0.3402\\\scriptsize{$(0.0011)$}} & \shortstack{0.3346\\\scriptsize{$(0.0027)$}} & \shortstack{0.3502\\\scriptsize{$(0.0018)$}} & \shortstack{\uline{0.3559}\\\scriptsize{$(0.0023)$}} & \shortstack{0.3380\\\scriptsize{$(0.0018)$}} & \shortstack{0.3381\\\scriptsize{$(0.0017)$}} & \shortstack{0.3438\\\scriptsize{$(0.0023)$}} & \shortstack{0.3037\\\scriptsize{$(0.0011)$}} & \shortstack{\textbf{0.3628}*\\\scriptsize{$(0.0025)$}} & \shortstack{3.60\%\\\;} & \shortstack{1.94\%\\\;}   \\
    & \shortstack{R@10\\\;} & \shortstack{0.2982\\\scriptsize{$(0.0014)$}} & \shortstack{0.2753\\\scriptsize{$(0.0019)$}} & \shortstack{0.3882\\\scriptsize{$(0.0026)$}} & \shortstack{0.3795\\\scriptsize{$(0.0019)$}} & \shortstack{0.3944\\\scriptsize{$(0.0029)$}} & \shortstack{0.3897\\\scriptsize{$(0.0037)$}} & \shortstack{0.4013\\\scriptsize{$(0.0022)$}} & \shortstack{\uline{0.4058}\\\scriptsize{$(0.0030)$}} & \shortstack{0.3939\\\scriptsize{$(0.0024)$}} & \shortstack{0.3931\\\scriptsize{$(0.0021)$}} & \shortstack{0.3986\\\scriptsize{$(0.0023)$}} & \shortstack{0.3648\\\scriptsize{$(0.0013)$}} & \shortstack{\textbf{0.4145}*\\\scriptsize{$(0.0029)$}} & \shortstack{3.29\%\\\;} & \shortstack{2.14\%\\\;}   \\
    & \shortstack{M@5\\\;} & \shortstack{0.0904\\\scriptsize{$(0.0009)$}} & \shortstack{0.1802\\\scriptsize{$(0.0021)$}} & \shortstack{0.2724\\\scriptsize{$(0.0018)$}} & \shortstack{0.2644\\\scriptsize{$(0.0014)$}} & \shortstack{0.2765\\\scriptsize{$(0.0021)$}} & \shortstack{0.2670\\\scriptsize{$(0.0016)$}} & \shortstack{0.2861\\\scriptsize{$(0.0019)$}} & \shortstack{\uline{0.2930}\\\scriptsize{$(0.0014)$}} & \shortstack{0.2689\\\scriptsize{$(0.0021)$}} & \shortstack{0.2695\\\scriptsize{$(0.0021)$}} & \shortstack{0.2758\\\scriptsize{$(0.0016)$}} & \shortstack{0.2324\\\scriptsize{$(0.0012)$}} & \shortstack{\textbf{0.2974}*\\\scriptsize{$(0.0020)$}} & \shortstack{3.95\%\\\;} & \shortstack{1.50\%\\\;}   \\
    & \shortstack{M@10\\\;} & \shortstack{0.1061\\\scriptsize{$(0.0009)$}} & \shortstack{0.1854\\\scriptsize{$(0.0020)$}} & \shortstack{0.2795\\\scriptsize{$(0.0018)$}} & \shortstack{0.2712\\\scriptsize{$(0.0014)$}} & \shortstack{0.2837\\\scriptsize{$(0.0019)$}} & \shortstack{0.2743\\\scriptsize{$(0.0016)$}} & \shortstack{0.2929\\\scriptsize{$(0.0019)$}} & \shortstack{\uline{0.2996}\\\scriptsize{$(0.0015)$}} & \shortstack{0.2763\\\scriptsize{$(0.0020)$}} & \shortstack{0.2768\\\scriptsize{$(0.0020)$}} & \shortstack{0.2831\\\scriptsize{$(0.0016)$}} & \shortstack{0.2404\\\scriptsize{$(0.0012)$}} & \shortstack{\textbf{0.3043}*\\\scriptsize{$(0.0020)$}} & \shortstack{3.89\%\\\;} & \shortstack{1.57\%\\\;}   \\
    & \shortstack{N@5\\\;} & \shortstack{0.1120\\\scriptsize{$(0.0012)$}} & \shortstack{0.1941\\\scriptsize{$(0.0020)$}} & \shortstack{0.2880\\\scriptsize{$(0.0018)$}} & \shortstack{0.2803\\\scriptsize{$(0.0014)$}} & \shortstack{0.2924\\\scriptsize{$(0.0016)$}} & \shortstack{0.2838\\\scriptsize{$(0.0018)$}} & \shortstack{0.3021\\\scriptsize{$(0.0019)$}} & \shortstack{\uline{0.3087}\\\scriptsize{$(0.0016)$}} & \shortstack{0.2861\\\scriptsize{$(0.0019)$}} & \shortstack{0.2866\\\scriptsize{$(0.0019)$}} & \shortstack{0.2927\\\scriptsize{$(0.0018)$}} & \shortstack{0.2501\\\scriptsize{$(0.0011)$}} & \shortstack{\textbf{0.3137}*\\\scriptsize{$(0.0021)$}} & \shortstack{3.84\%\\\;} & \shortstack{1.62\%\\\;}   \\
    & \shortstack{N@10\\\;} & \shortstack{0.1505\\\scriptsize{$(0.0010)$}} & \shortstack{0.2068\\\scriptsize{$(0.0020)$}} & \shortstack{0.3052\\\scriptsize{$(0.0018)$}} & \shortstack{0.2968\\\scriptsize{$(0.0015)$}} & \shortstack{0.3099\\\scriptsize{$(0.0012)$}} & \shortstack{0.3016\\\scriptsize{$(0.0019)$}} & \shortstack{0.3186\\\scriptsize{$(0.0018)$}} & \shortstack{\uline{0.3248}\\\scriptsize{$(0.0019)$}} & \shortstack{0.3041\\\scriptsize{$(0.0019)$}} & \shortstack{0.3043\\\scriptsize{$(0.0017)$}} & \shortstack{0.3104\\\scriptsize{$(0.0018)$}} & \shortstack{0.2698\\\scriptsize{$(0.0011)$}} & \shortstack{\textbf{0.3304}*\\\scriptsize{$(0.0022)$}} & \shortstack{3.70\%\\\;} & \shortstack{1.72\%\\\;}   \\
    \midrule
\multirow{6}{*}{\begin{tabular}[c]{@{}c@{}}MSSD\\ 5d\end{tabular}}
    & \shortstack{R@5\\\;} & \shortstack{0.1712\\\scriptsize{$(0.0009)$}} & \shortstack{0.2329\\\scriptsize{$(0.0010)$}} & \shortstack{0.3394\\\scriptsize{$(0.0016)$}} & \shortstack{0.3316\\\scriptsize{$(0.0014)$}} & \shortstack{0.3440\\\scriptsize{$(0.0013)$}} & \shortstack{0.3350\\\scriptsize{$(0.0017)$}} & \shortstack{0.3529\\\scriptsize{$(0.0010)$}} & \shortstack{\uline{0.3562}\\\scriptsize{$(0.0012)$}} & \shortstack{0.3352\\\scriptsize{$(0.0016)$}} & \shortstack{0.3378\\\scriptsize{$(0.0020)$}} & \shortstack{0.3438\\\scriptsize{$(0.0012)$}} & \shortstack{0.3159\\\scriptsize{$(0.0020)$}} & \shortstack{\textbf{0.3636}*\\\scriptsize{$(0.0005)$}} & \shortstack{3.03\%\\\;} & \shortstack{2.08\%\\\;}   \\
    & \shortstack{R@10\\\;} & \shortstack{0.2884\\\scriptsize{$(0.0013)$}} & \shortstack{0.2745\\\scriptsize{$(0.0014)$}} & \shortstack{0.3941\\\scriptsize{$(0.0032)$}} & \shortstack{0.3841\\\scriptsize{$(0.0010)$}} & \shortstack{0.3990\\\scriptsize{$(0.0018)$}} & \shortstack{0.3891\\\scriptsize{$(0.0021)$}} & \shortstack{0.4040\\\scriptsize{$(0.0020)$}} & \shortstack{\uline{0.4065}\\\scriptsize{$(0.0015)$}} & \shortstack{0.3886\\\scriptsize{$(0.0019)$}} & \shortstack{0.3926\\\scriptsize{$(0.0026)$}} & \shortstack{0.3989\\\scriptsize{$(0.0015)$}} & \shortstack{0.3747\\\scriptsize{$(0.0012)$}} & \shortstack{\textbf{0.4153}*\\\scriptsize{$(0.0008)$}} & \shortstack{2.80\%\\\;} & \shortstack{2.16\%\\\;}   \\
    & \shortstack{M@5\\\;} & \shortstack{0.0872\\\scriptsize{$(0.0006)$}} & \shortstack{0.1745\\\scriptsize{$(0.0004)$}} & \shortstack{0.2764\\\scriptsize{$(0.0005)$}} & \shortstack{0.2671\\\scriptsize{$(0.0014)$}} & \shortstack{0.2804\\\scriptsize{$(0.0013)$}} & \shortstack{0.2701\\\scriptsize{$(0.0014)$}} & \shortstack{0.2899\\\scriptsize{$(0.0007)$}} & \shortstack{\uline{0.2939}\\\scriptsize{$(0.0011)$}} & \shortstack{0.2711\\\scriptsize{$(0.0010)$}} & \shortstack{0.2717\\\scriptsize{$(0.0015)$}} & \shortstack{0.2769\\\scriptsize{$(0.0013)$}} & \shortstack{0.2476\\\scriptsize{$(0.0023)$}} & \shortstack{\textbf{0.2993}*\\\scriptsize{$(0.0006)$}} & \shortstack{3.24\%\\\;} & \shortstack{1.84\%\\\;}   \\
    & \shortstack{M@10\\\;} & \shortstack{0.1025\\\scriptsize{$(0.0006)$}} & \shortstack{0.1800\\\scriptsize{$(0.0005)$}} & \shortstack{0.2836\\\scriptsize{$(0.0007)$}} & \shortstack{0.2741\\\scriptsize{$(0.0013)$}} & \shortstack{0.2876\\\scriptsize{$(0.0013)$}} & \shortstack{0.2772\\\scriptsize{$(0.0013)$}} & \shortstack{0.2967\\\scriptsize{$(0.0007)$}} & \shortstack{\uline{0.3006}\\\scriptsize{$(0.0011)$}} & \shortstack{0.2781\\\scriptsize{$(0.0010)$}} & \shortstack{0.2790\\\scriptsize{$(0.0015)$}} & \shortstack{0.2843\\\scriptsize{$(0.0011)$}} & \shortstack{0.2554\\\scriptsize{$(0.0022)$}} & \shortstack{\textbf{0.3062}*\\\scriptsize{$(0.0005)$}} & \shortstack{3.20\%\\\;} & \shortstack{1.86\%\\\;}   \\
    & \shortstack{N@5\\\;} & \shortstack{0.1078\\\scriptsize{$(0.0006)$}} & \shortstack{0.1890\\\scriptsize{$(0.0005)$}} & \shortstack{0.2920\\\scriptsize{$(0.0008)$}} & \shortstack{0.2832\\\scriptsize{$(0.0013)$}} & \shortstack{0.2962\\\scriptsize{$(0.0012)$}} & \shortstack{0.2863\\\scriptsize{$(0.0014)$}} & \shortstack{0.3056\\\scriptsize{$(0.0006)$}} & \shortstack{\uline{0.3094}\\\scriptsize{$(0.0011)$}} & \shortstack{0.2870\\\scriptsize{$(0.0011)$}} & \shortstack{0.2882\\\scriptsize{$(0.0016)$}} & \shortstack{0.2936\\\scriptsize{$(0.0011)$}} & \shortstack{0.2646\\\scriptsize{$(0.0022)$}} & \shortstack{\textbf{0.3154}*\\\scriptsize{$(0.0005)$}} & \shortstack{3.21\%\\\;} & \shortstack{1.94\%\\\;}   \\
    & \shortstack{N@10\\\;} & \shortstack{0.1455\\\scriptsize{$(0.0007)$}} & \shortstack{0.2025\\\scriptsize{$(0.0007)$}} & \shortstack{0.3096\\\scriptsize{$(0.0012)$}} & \shortstack{0.3001\\\scriptsize{$(0.0012)$}} & \shortstack{0.3139\\\scriptsize{$(0.0012)$}} & \shortstack{0.3037\\\scriptsize{$(0.0013)$}} & \shortstack{0.3221\\\scriptsize{$(0.0008)$}} & \shortstack{\uline{0.3257}\\\scriptsize{$(0.0011)$}} & \shortstack{0.3042\\\scriptsize{$(0.0011)$}} & \shortstack{0.3059\\\scriptsize{$(0.0018)$}} & \shortstack{0.3114\\\scriptsize{$(0.0010)$}} & \shortstack{0.2836\\\scriptsize{$(0.0019)$}} & \shortstack{\textbf{0.3320}*\\\scriptsize{$(0.0004)$}} & \shortstack{3.07\%\\\;} & \shortstack{1.93\%\\\;}   \\
    \midrule
\multirow{6}{*}{\begin{tabular}[c]{@{}c@{}}MSSD\\ 7d\end{tabular}}
    & \shortstack{R@5\\\;} & \shortstack{0.1749\\\scriptsize{$(0.0020)$}} & \shortstack{0.2259\\\scriptsize{$(0.0019)$}} & \shortstack{0.3363\\\scriptsize{$(0.0013)$}} & \shortstack{0.3257\\\scriptsize{$(0.0007)$}} & \shortstack{0.3388\\\scriptsize{$(0.0009)$}} & \shortstack{0.3314\\\scriptsize{$(0.0019)$}} & \shortstack{0.3498\\\scriptsize{$(0.0014)$}} & \shortstack{\uline{0.3522}\\\scriptsize{$(0.0013)$}} & \shortstack{0.3336\\\scriptsize{$(0.0011)$}} & \shortstack{0.3344\\\scriptsize{$(0.0009)$}} & \shortstack{0.3401\\\scriptsize{$(0.0017)$}} & \shortstack{0.3086\\\scriptsize{$(0.0031)$}} & \shortstack{\textbf{0.3607}*\\\scriptsize{$(0.0018)$}} & \shortstack{3.12\%\\\;} & \shortstack{2.41\%\\\;}   \\
    & \shortstack{R@10\\\;} & \shortstack{0.2903\\\scriptsize{$(0.0018)$}} & \shortstack{0.2699\\\scriptsize{$(0.0020)$}} & \shortstack{0.3943\\\scriptsize{$(0.0022)$}} & \shortstack{0.3803\\\scriptsize{$(0.0006)$}} & \shortstack{0.3960\\\scriptsize{$(0.0012)$}} & \shortstack{0.3887\\\scriptsize{$(0.0029)$}} & \shortstack{0.4038\\\scriptsize{$(0.0016)$}} & \shortstack{\uline{0.4054}\\\scriptsize{$(0.0016)$}} & \shortstack{0.3906\\\scriptsize{$(0.0012)$}} & \shortstack{0.3917\\\scriptsize{$(0.0009)$}} & \shortstack{0.3979\\\scriptsize{$(0.0020)$}} & \shortstack{0.3695\\\scriptsize{$(0.0028)$}} & \shortstack{\textbf{0.4150}*\\\scriptsize{$(0.0024)$}} & \shortstack{2.77\%\\\;} & \shortstack{2.37\%\\\;}   \\
    & \shortstack{M@5\\\;} & \shortstack{0.0898\\\scriptsize{$(0.0011)$}} & \shortstack{0.1653\\\scriptsize{$(0.0018)$}} & \shortstack{0.2690\\\scriptsize{$(0.0006)$}} & \shortstack{0.2584\\\scriptsize{$(0.0011)$}} & \shortstack{0.2732\\\scriptsize{$(0.0006)$}} & \shortstack{0.2625\\\scriptsize{$(0.0008)$}} & \shortstack{0.2832\\\scriptsize{$(0.0006)$}} & \shortstack{\uline{0.2860}\\\scriptsize{$(0.0007)$}} & \shortstack{0.2647\\\scriptsize{$(0.0012)$}} & \shortstack{0.2650\\\scriptsize{$(0.0008)$}} & \shortstack{0.2698\\\scriptsize{$(0.0011)$}} & \shortstack{0.2394\\\scriptsize{$(0.0030)$}} & \shortstack{\textbf{0.2929}*\\\scriptsize{$(0.0012)$}} & \shortstack{3.43\%\\\;} & \shortstack{2.41\%\\\;}   \\
    & \shortstack{M@10\\\;} & \shortstack{0.1049\\\scriptsize{$(0.0011)$}} & \shortstack{0.1711\\\scriptsize{$(0.0018)$}} & \shortstack{0.2766\\\scriptsize{$(0.0006)$}} & \shortstack{0.2656\\\scriptsize{$(0.0010)$}} & \shortstack{0.2807\\\scriptsize{$(0.0006)$}} & \shortstack{0.2701\\\scriptsize{$(0.0008)$}} & \shortstack{0.2904\\\scriptsize{$(0.0007)$}} & \shortstack{\uline{0.2931}\\\scriptsize{$(0.0006)$}} & \shortstack{0.2722\\\scriptsize{$(0.0012)$}} & \shortstack{0.2725\\\scriptsize{$(0.0008)$}} & \shortstack{0.2775\\\scriptsize{$(0.0011)$}} & \shortstack{0.2474\\\scriptsize{$(0.0030)$}} & \shortstack{\textbf{0.3002}*\\\scriptsize{$(0.0012)$}} & \shortstack{3.37\%\\\;} & \shortstack{2.42\%\\\;}   \\
    & \shortstack{N@5\\\;} & \shortstack{0.1106\\\scriptsize{$(0.0013)$}} & \shortstack{0.1804\\\scriptsize{$(0.0018)$}} & \shortstack{0.2857\\\scriptsize{$(0.0007)$}} & \shortstack{0.2751\\\scriptsize{$(0.0009)$}} & \shortstack{0.2895\\\scriptsize{$(0.0006)$}} & \shortstack{0.2796\\\scriptsize{$(0.0009)$}} & \shortstack{0.2999\\\scriptsize{$(0.0008)$}} & \shortstack{\uline{0.3025}\\\scriptsize{$(0.0007)$}} & \shortstack{0.2818\\\scriptsize{$(0.0012)$}} & \shortstack{0.2822\\\scriptsize{$(0.0008)$}} & \shortstack{0.2873\\\scriptsize{$(0.0012)$}} & \shortstack{0.2566\\\scriptsize{$(0.0030)$}} & \shortstack{\textbf{0.3099}*\\\scriptsize{$(0.0013)$}} & \shortstack{3.33\%\\\;} & \shortstack{2.45\%\\\;}   \\
    & \shortstack{N@10\\\;} & \shortstack{0.1478\\\scriptsize{$(0.0012)$}} & \shortstack{0.1945\\\scriptsize{$(0.0018)$}} & \shortstack{0.3044\\\scriptsize{$(0.0009)$}} & \shortstack{0.2928\\\scriptsize{$(0.0008)$}} & \shortstack{0.3079\\\scriptsize{$(0.0007)$}} & \shortstack{0.2981\\\scriptsize{$(0.0010)$}} & \shortstack{0.3172\\\scriptsize{$(0.0009)$}} & \shortstack{\uline{0.3197}\\\scriptsize{$(0.0007)$}} & \shortstack{0.3002\\\scriptsize{$(0.0011)$}} & \shortstack{0.3007\\\scriptsize{$(0.0009)$}} & \shortstack{0.3059\\\scriptsize{$(0.0012)$}} & \shortstack{0.2762\\\scriptsize{$(0.0029)$}} & \shortstack{\textbf{0.3274}*\\\scriptsize{$(0.0013)$}} & \shortstack{3.22\%\\\;} & \shortstack{2.41\%\\\;}   \\
    \bottomrule
\end{tabular}
}
\vspace{-3ex}
\label{tab:performance}
\end{table*}

\subsection{Overall Performance Comparison}
To demonstrate the effectiveness of \proposed, we report the recommendation accuracy of \proposed~and all the baselines in Table~\ref{tab:performance}. \proposed~ aims to fully leverage the shuffle play sessions through transition-based augmentation and fine-grained matching strategies, i.e., item- and similarity-based matching.
We summarize the following observations:
1) \proposed~achieves state-of-the-art performance in the real-world, large-scale dataset (i.e., MSSD) over 12 baseline recommender systems, which demonstrates the superiority of \proposed.
2) CoSeRNN, a music recommender system, performs inferior to \proposed{} and other baseline models. We speculate that CoSeRNN is designed to depend heavily on user identity and contextual information (e.g., device type), while they are not provided in MSSD. 
In contrast, \proposed~ shows superior performance without the auxiliary information, which signifies the practicality of \proposed.
3) Graph-based methods, e.g., SRGNN and GCSAN, show the highest performance over the other baseline methods. The graph-based methods mainly utilize the transition between tracks in sessions by constructing graphs. Thus, it implies that transition information is important in music recommendation. \proposed~ also takes the ability of graph by taking SRGNN as the backbone. In addition to GNN, the transition-based augmentation further supplements the transition information into the shuffle-play sessions, which supports the superior performance of \proposed. 
4) As self-supervised learning (SSL) approaches, e.g., CL4SRec and DuoRec, improve the performance of the backbone (i.e., SASRec), \proposed~significantly outperforms SRGNN. It implies that the SSL framework fully utilizes the backbone's ability with the same number of parameters.
5) Additionally, \proposed~significantly surpasses other SSL approaches. It is shown that our SSL frameworks and fine-grained matching strategies facilitate the alignment of representations.

\begin{table}[t]
\caption[Performance on shuffle play sessions]{Performance on shuffle play sessions.}
\vspace{-3ex}
\resizebox{0.95\columnwidth}{!}{
\begin{tabular}{@{}clcccccccc@{}}
\toprule
\multicolumn{2}{c}{Setting}                                                                     & Rec. SBR    & SSL SBR & \multicolumn{2}{c}{Graph-based SBR} & Ours          & \multicolumn{3}{c}{Relative Gap} \\ 
\cmidrule(lr){1-2} \cmidrule(lr){3-3} \cmidrule(lr){4-4} \cmidrule(lr){5-6} \cmidrule(lr){7-7} \cmidrule(lr){8-10}
Dataset                                                            & \multicolumn{1}{c}{Metric} & FMLP          & CL4SRec       & SRGNN            & GCSAN            & \textbf{\proposed}        & $\Delta_b$  & $\Delta_s$ \\ \midrule
\multirow{4}{*}{\begin{tabular}[c]{@{}c@{}}MSSD\\ 3d\end{tabular}}
    & \shortstack{R@10\\\;} & \shortstack{0.2256\\\scriptsize{$(0.0009)$}} & \shortstack{0.2297\\\scriptsize{$(0.0025)$}} & \shortstack{\uline{0.2304}\\\scriptsize{$(0.0024)$}} & \shortstack{0.2283\\\scriptsize{$(0.0020)$}} & \shortstack{\textbf{0.2401}*\\\scriptsize{$(0.0015)$}} & \shortstack{4.21\%\\\;} & \shortstack{5.17\%\\\;}   \\
    & \shortstack{M@10\\\;} & \shortstack{0.1071\\\scriptsize{$(0.0008)$}} & \shortstack{0.1080\\\scriptsize{$(0.0014)$}} & \shortstack{\uline{0.1140}\\\scriptsize{$(0.0010)$}} & \shortstack{0.1137\\\scriptsize{$(0.0013)$}} & \shortstack{\textbf{0.1181}*\\\scriptsize{$(0.0008)$}} & \shortstack{3.60\%\\\;} & \shortstack{3.87\%\\\;}   \\
    & \shortstack{N@10\\\;} & \shortstack{0.1345\\\scriptsize{$(0.0007)$}} & \shortstack{0.1362\\\scriptsize{$(0.0016)$}} & \shortstack{\uline{0.1410}\\\scriptsize{$(0.0013)$}} & \shortstack{0.1402\\\scriptsize{$(0.0014)$}} & \shortstack{\textbf{0.1464}*\\\scriptsize{$(0.0009)$}} & \shortstack{3.83\%\\\;} & \shortstack{4.42\%\\\;}   \\
    \midrule
\multirow{4}{*}{\begin{tabular}[c]{@{}c@{}}MSSD\\ 5d\end{tabular}}
    & \shortstack{R@10\\\;} & \shortstack{0.2265\\\scriptsize{$(0.0011)$}} & \shortstack{0.2250\\\scriptsize{$(0.0015)$}} & \shortstack{\uline{0.2330}\\\scriptsize{$(0.0023)$}} & \shortstack{0.2295\\\scriptsize{$(0.0017)$}} & \shortstack{\textbf{0.2400}*\\\scriptsize{$(0.0012)$}} & \shortstack{3.00\%\\\;} & \shortstack{4.58\%\\\;}   \\
    & \shortstack{M@10\\\;} & \shortstack{0.1069\\\scriptsize{$(0.0010)$}} & \shortstack{0.1061\\\scriptsize{$(0.0008)$}} & \shortstack{\uline{0.1146}\\\scriptsize{$(0.0010)$}} & \shortstack{0.1136\\\scriptsize{$(0.0010)$}} & \shortstack{\textbf{0.1179}*\\\scriptsize{$(0.0004)$}} & \shortstack{2.88\%\\\;} & \shortstack{3.79\%\\\;}   \\
    & \shortstack{N@10\\\;} & \shortstack{0.1345\\\scriptsize{$(0.0008)$}} & \shortstack{0.1337\\\scriptsize{$(0.0007)$}} & \shortstack{\uline{0.1420}\\\scriptsize{$(0.0011)$}} & \shortstack{0.1404\\\scriptsize{$(0.0010)$}} & \shortstack{\textbf{0.1462}*\\\scriptsize{$(0.0003)$}} & \shortstack{2.96\%\\\;} & \shortstack{4.13\%\\\;}   \\
    \bottomrule
\end{tabular}
\label{tab:shuffle}
}
\vspace{-3.5ex}
\end{table}

\begin{table}[t]
\caption[Performance on non-shuffle play sessions]{Performance on non-shuffle play sessions.}
\vspace{-3ex}
\resizebox{0.95\columnwidth}{!}{
\begin{tabular}{@{}clcccccccc@{}}
\toprule
\multicolumn{2}{c}{Setting}                                                                     & Rec. SBR    & SSL SBR & \multicolumn{2}{c}{Graph-based SBR} & Ours          & \multicolumn{3}{c}{Relative Gap} \\ \cmidrule(lr){1-2} \cmidrule(lr){3-3} \cmidrule(lr){4-4} \cmidrule(lr){5-6} \cmidrule(lr){7-7} \cmidrule(lr){8-10}
Dataset                                                            & \multicolumn{1}{c}{Metric} & FMLP          & CL4SRec       & SRGNN            & GCSAN            & \textbf{\proposed}      & $\Delta_b$  & $\Delta_s$ \\ \midrule
\multirow{4}{*}{\begin{tabular}[c]{@{}c@{}}MSSD\\ 3d\end{tabular}}
    & \shortstack{R@10\\\;} & \shortstack{0.4868\\\scriptsize{$(0.0032)$}} & \shortstack{0.4776\\\scriptsize{$(0.0029)$}} & \shortstack{0.4885\\\scriptsize{$(0.0029)$}} & \shortstack{\uline{0.4963}\\\scriptsize{$(0.0038)$}} & \shortstack{\textbf{0.5034}*\\\scriptsize{$(0.0037)$}} & \shortstack{3.05\%\\\;} & \shortstack{1.43\%\\\;}   \\
    & \shortstack{M@10\\\;} & \shortstack{0.3728\\\scriptsize{$(0.0026)$}} & \shortstack{0.3620\\\scriptsize{$(0.0032)$}} & \shortstack{0.3841\\\scriptsize{$(0.0034)$}} & \shortstack{\uline{0.3943}\\\scriptsize{$(0.0024)$}} & \shortstack{\textbf{0.3992}*\\\scriptsize{$(0.0030)$}} & \shortstack{3.93\%\\\;} & \shortstack{1.24\%\\\;}   \\
    & \shortstack{N@10\\\;} & \shortstack{0.4001\\\scriptsize{$(0.0028)$}} & \shortstack{0.3897\\\scriptsize{$(0.0028)$}} & \shortstack{0.4091\\\scriptsize{$(0.0031)$}} & \shortstack{\uline{0.4188}\\\scriptsize{$(0.0026)$}} & \shortstack{\textbf{0.4242}*\\\scriptsize{$(0.0031)$}} & \shortstack{3.69\%\\\;} & \shortstack{1.29\%\\\;}   \\
    \midrule
\multirow{4}{*}{\begin{tabular}[c]{@{}c@{}}MSSD\\ 5d\end{tabular}}
    & \shortstack{R@10\\\;} & \shortstack{0.4872\\\scriptsize{$(0.0017)$}} & \shortstack{0.4724\\\scriptsize{$(0.0021)$}} & \shortstack{0.4916\\\scriptsize{$(0.0025)$}} & \shortstack{\uline{0.4972}\\\scriptsize{$(0.0014)$}} & \shortstack{\textbf{0.5051}*\\\scriptsize{$(0.0007)$}} & \shortstack{2.75\%\\\;} & \shortstack{1.59\%\\\;}   \\
    & \shortstack{M@10\\\;} & \shortstack{0.3751\\\scriptsize{$(0.0006)$}} & \shortstack{0.3662\\\scriptsize{$(0.0008)$}} & \shortstack{0.3899\\\scriptsize{$(0.0007)$}} & \shortstack{\uline{0.3963}\\\scriptsize{$(0.0008)$}} & \shortstack{\textbf{0.4026}*\\\scriptsize{$(0.0008)$}} & \shortstack{3.26\%\\\;} & \shortstack{1.59\%\\\;}   \\
    & \shortstack{N@10\\\;} & \shortstack{0.4019\\\scriptsize{$(0.0005)$}} & \shortstack{0.3916\\\scriptsize{$(0.0011)$}} & \shortstack{0.4143\\\scriptsize{$(0.0010)$}} & \shortstack{\uline{0.4205}\\\scriptsize{$(0.0010)$}} & \shortstack{\textbf{0.4272}*\\\scriptsize{$(0.0007)$}} & \shortstack{3.11\%\\\;} & \shortstack{1.59\%\\\;}   \\
    \bottomrule
\end{tabular}
\label{tab:nonshuffle}
}
\vspace{-5ex}
\end{table}

\subsection{Fine-grained Performance Comparison}
We delve into the examination of \proposed~ on fine-grained scenarios to deeply understand its benefits.
We divide the test data into two subsets consisting of the shuffle and non-shuffle play sessions in Table~\ref{tab:shuffle} and Table~\ref{tab:nonshuffle}, respectively. We make the following observations:
1) \proposed~substantially bolsters the performance on the shuffle play sessions compared to baselines (see Table \ref{tab:shuffle}). It indicates that the transition-based augmentation and fine-grained matching strategies of \proposed~ are indeed beneficial to shuffle play sessions. It aligns with our primary objective of improving shuffle-play sessions by alleviating the unique transitions. 
2) \proposed~ also boosts the performance on non-shuffle play sessions even though our framework focuses on shuffle play sessions (see Table \ref{tab:nonshuffle}). The SSL framework can enhance the representation of non-shuffle play sessions by utilizing reorder-based augmentation.
{3) The performance gain of the state-of-the-art baseline, GCSAN, is biased towards non-shuffle play sessions (see Table \ref{tab:nonshuffle}), as it struggles to surpass its backbone, SRGNN, during shuffle play sessions (see Table \ref{tab:shuffle}). This indicates that self-attention falls short of capturing users' dynamic preferences in the shuffle play environment.} In summary,~\proposed~demonstrates its efficacy in improving not only the performance of shuffle play sessions but also non-shuffle play sessions. {This is achieved by addressing unique transitions through transition-based augmentation, enhancing robustness via fine-grained matching between augmented views, and mimicking shuffle play environment using reorder-based augmentation for non-shuffle play sessions.} It highlights the versatility and effectiveness of~\proposed~in capturing user preferences across diverse music playback sessions.

\begin{figure}[!t]
    \includegraphics[width=1\columnwidth]{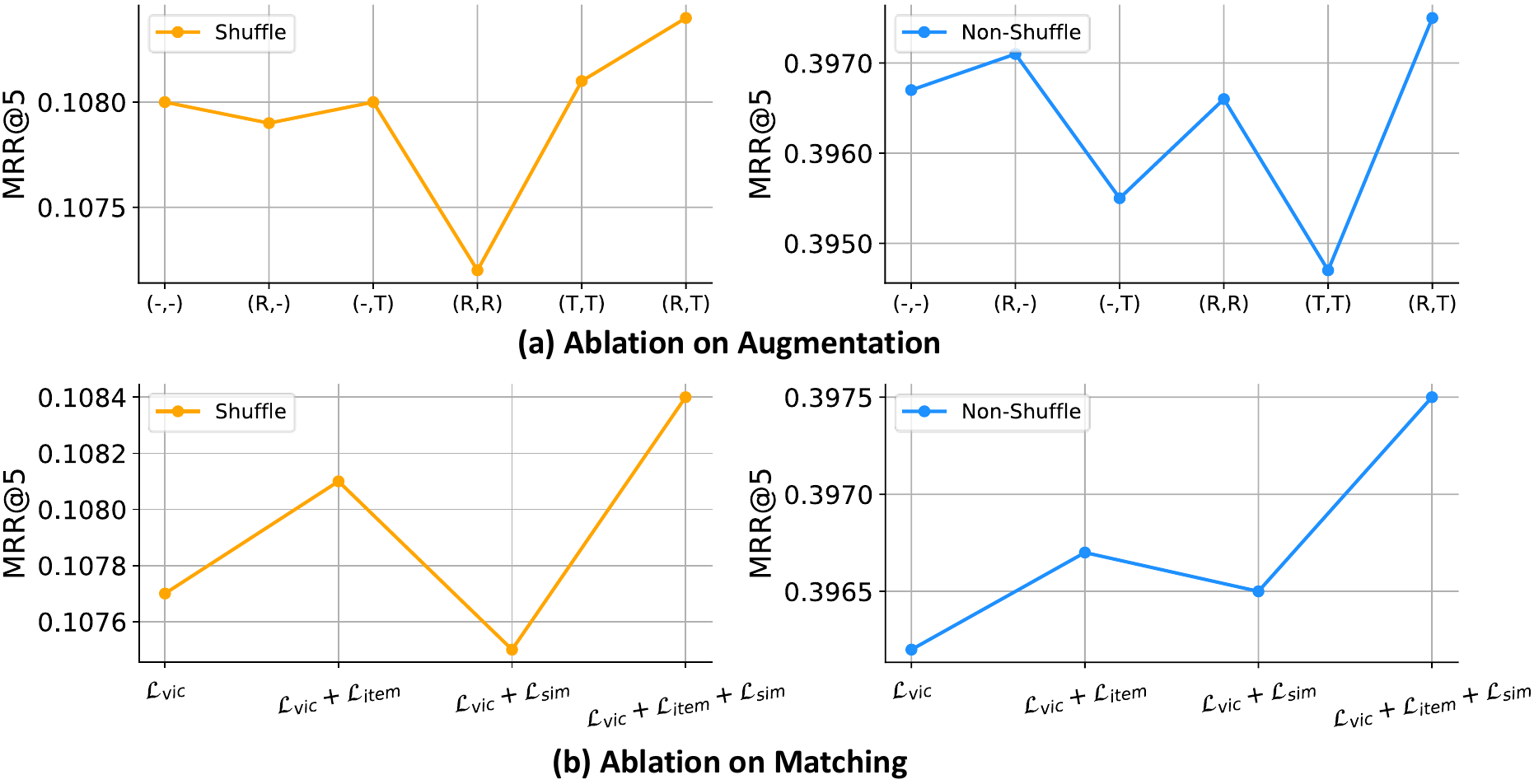}
    \vspace{-5ex}
    \caption[Ablation studies]{\label{fig:ablation} Ablation studies (MSSD-5d dataset, MRR@5). In (a), "(-,-)" denotes when augmentation is not made on both non-shuffle (left) and shuffle play sessions (right), "(R,-)", "(-,T)" denote reorder-based augmentation is made on the non-shuffle play sessions, and transition-based augmentation is made on shuffle play sessions, respectively. (R,T):~\proposed.}
    \vspace{-3ex}
\end{figure}

\subsection{Ablation Studies}
\noindent{\textbf{Ablation on Transition-based Augmentation.}}
In Figure~\ref{fig:ablation} (a), our observations suggest that applying augmentations to both shuffle and non-shuffle play sessions yields optimal results. More specifically, non-shuffle play sessions benefit from reorder-based augmentation, while shuffle play sessions derive particular advantages from transition-based augmentation (i.e., (R,T) yields the best result). These findings underscore the significance of incorporating transition information for shuffle play sessions and highlight the efficacy of reordering tracks for non-shuffle play sessions.

\noindent{\textbf{Ablation on Fine-grained Matching.}}
In Figure \ref{fig:ablation} (b), we present an ablation study to delve into the effectiveness of the fine-grained matching strategies. These strategies demonstrate enhancements in recommendations for both shuffle and non-shuffle play sessions. More precisely, item-based matching facilitates the alignment of the track embeddings of the identical items between two views. It enables us to cope with the shuffle and non-shuffle play session recommendations adeptly. As elaborated in Section 3.3, similarity-based matching complements item-based matching by considering the similarity of track representations. 
This synergistic combination verifies that the similarity-based method reinforces the item-based approach as envisioned, leading to a more refined and precise alignment process. In essence, employing both item-based and similarity-based matching mechanisms is pivotal in optimizing recommendation performance.

\begin{figure}[!t]
    \includegraphics[width=.7\columnwidth]{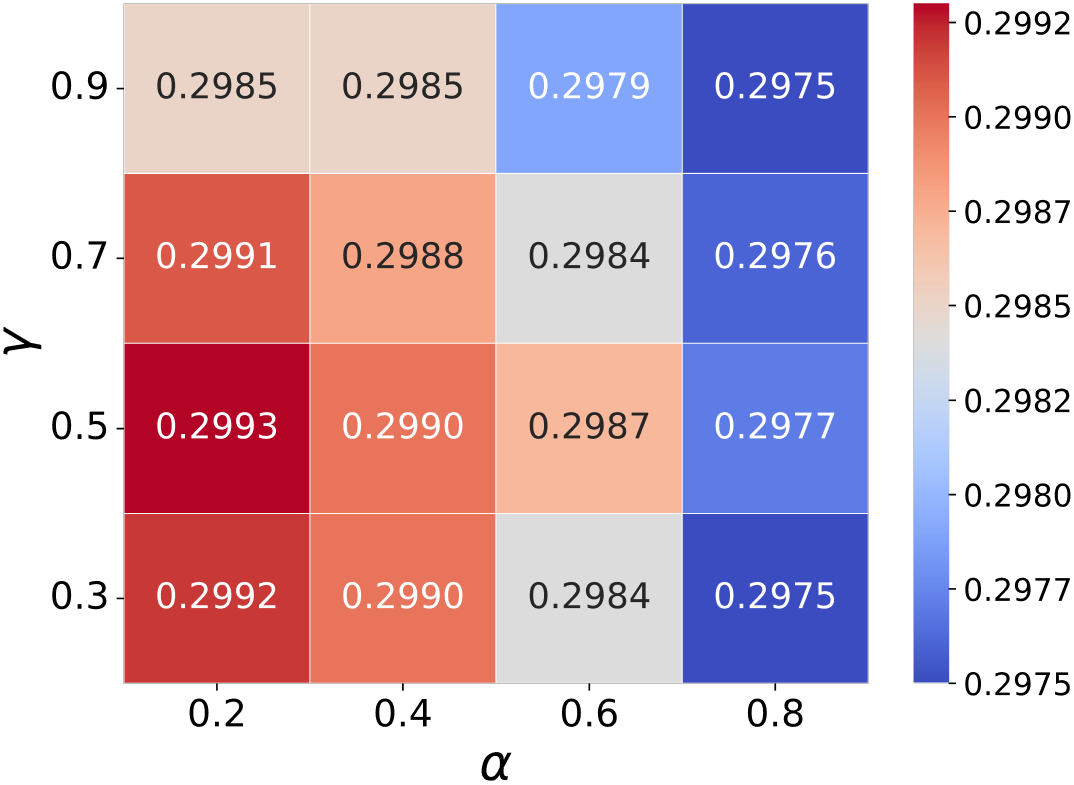}
    \vspace{-3ex}
    \caption[Sensitivity analysis]{\label{fig:sensitivity} Sensitivity analysis (MSSD-5d dataset, MRR@5). $\alpha$ and $\gamma$ are loss-controlling and reordering hyperparameters, respectively.}
    \vspace{-3ex}
\end{figure}

\subsection{Sensitivity Analysis}
The two key hyperparameters for~\proposed~are reordering hyperparameter $\gamma$, which determines the proportion of tracks to be reordered, and loss-controlling hyperparameter $\alpha$ (Eq. \ref{loss:final}), which balances between the matching loss and alignment loss. {As depicted in Figure~\ref{fig:sensitivity},~\proposed~demonstrates robust performance, outperforming the state-of-the-art baseline, GCSAN~\cite{gcsan} (0.2939 as shown in Table~\ref{tab:performance}), in all combinations. We notice that a moderate reordering probability, $\gamma$ (i.e., 0.5), of non-shuffle play sessions is advantageous. Excessive reordering (i.e., high $\gamma$) could hamper the original session's semantics, while too little reordering (i.e., low $\gamma$) might hinder the augmentation's potential for enhancing generalizability. This result aligns with our strategy, aiming to simulate the shuffle play context and enhance the model's capability to handle such scenarios effectively. Furthermore, opting for a low value of the loss-controlling hyperparameter, $\alpha$ (i.e., 0.2), proves to be advantageous for training \proposed. This is because $\mathcal{L}_{matching}$ encompasses both our proposed item- and similarity-based matching, leading to a relatively high scale of loss value. As a result, this choice acts effectively as a regularizer, contributing to the overall performance.

\section{Conclusion}
In this work, based on our empirical findings of the importance of shuffle play sessions in the music domain, we propose \proposed, a pioneering framework for music recommendation. Our approach employs a self-supervised learning framework to maximize the agreement between the original and augmented sessions. The augmentation is derived from a novel augmentation called transition-based augmentation, which alleviates the unique transition problem observed in shuffle play sessions by inserting the potential transition patterns. To further facilitate the alignment of representations across the two views, we introduce two precise matching strategies: the item-based approach ensuring proximity in the embedding space for identical items across both views, and the similarity-based matching strategy, which supplements the alignment of similar embeddings between the views based on the nearest neighbors of each track. Through experiments conducted across diverse environments, we demonstrate \proposed’s competence, specifically in the shuffle play environment, over 12 baseline models on a large-scale Music Streaming Sessions Dataset (MSSD) from Spotify. 
Moreover, a detailed analysis not only confirms \proposed's effectiveness in elevating performance for shuffle play sessions but also underscores its ability to bolster outcomes in non-shuffle play environments.

\smallskip
\looseness=-1
\noindent\textbf{Acknowledgement}: This work was supported by the National Research Foundation of Korea(NRF) grant funded by the Korea government(MSIT) (No.2021R1C1C1009081) and Institute of Information \& communications Technology Planning \& Evaluation (IITP) grant funded by the Korea government(MSIT) (No.2022-0-00077).

\bibliographystyle{ACM-Reference-Format}
\bibliography{MUSE}

\end{document}